\documentclass[aps,twocolumn,showpacs,preprintnumbers,amsmath,amssymb,nofootinbib,prc,superscriptaddress]{revtex4-1}

\usepackage{graphicx}
\usepackage[caption=false]{subfig}
\usepackage{amsmath}
\usepackage{mathtools}
\usepackage{float}
\usepackage{epstopdf}
\usepackage{color}
\usepackage{enumerate}

\newcommand{\beq}{\begin{equation}}
\newcommand{\enq}{\end{equation}}
\newcommand{\beqn}{\begin{eqnarray}}
\newcommand{\enqn}{\end{eqnarray}}
\newcommand{\al}{\alpha}
\newcommand{\be}{\beta}
\newcommand{\ga}{\gamma}
\newcommand{\de}{\delta}
\newcommand{\ep}{\epsilon}
\newcommand{\lm}{\lambda}
\newcommand{\om}{\omega}
\newcommand{\vep}{\varepsilon}
\newcommand{\ta}{\theta}
\newcommand{\sig}{\sigma}
\def\t{\tau}
\newcommand{\gz}{G^{(0)}}
\newcommand{\f}[2]{\frac{#1}{#2}}
\renewcommand{\d}{\mathrm d}
\newcommand{\lphizero}{\langle\Phi_0^N|}
\newcommand{\rphizero}{|\Phi_0^N\rangle}
\newcommand{\h}[1]{\hat{#1}}
\newcommand{\dg}{\dagger}
\newcommand{\T}{{\cal T}}
\newcommand{\nn}{\nonumber}
\newcommand{\dev}{\frac{\partial}{\partial t}}
\renewcommand{\a}[2]{a_{#1#2}}
\newcommand{\ad}[1]{a_{#1}^\dagger}
\newcommand{\ii}{\textrm{i}}
\def\ran{\rangle}
\def\lan{\langle}

\begin{document}

\title{Self-consistent Green's functions formalism with three-body interactions}

\author{Arianna Carbone}
\affiliation{ Departament d'Estructura i Constituents de la Mat\`eria and Institut de 
Ci\`{e}ncies del Cosmos, Universitat de Barcelona, E-08028 Barcelona, Spain }

\author{Andrea Cipollone}
\author{Carlo Barbieri}
\author{Arnau Rios}
\affiliation{ Department of Physics, Faculty of Engineering and Physical Sciences,
University of Surrey, Guildford, Surrey GU2 7XH, UK }

\author{Artur Polls}
\affiliation{ Departament d'Estructura i Constituents de la Mat\`eria and Institut de 
Ci\`{e}ncies del Cosmos, Universitat de Barcelona, E-08028 Barcelona, Spain }

\date{\today}

\begin{abstract}
We extend the self-consistent Green's functions formalism to take into account three-body interactions.
We analyze the perturbative expansion in terms of Feynman diagrams and define effective one- and two-body interactions, which allows for a substantial reduction of the 
number of diagrams.
The procedure can be taken as a generalization of 
the normal ordering of the Hamiltonian to fully correlated density matrices. 
We give examples up to third order 
in perturbation theory. To define nonperturbative approximations, we 
extend the equation of motion method in the presence of three-body interactions.
We propose schemes that can provide nonperturbative resummation of 
three-body interactions.
We also discuss two different extensions of the Koltun sum rule to compute the
ground state of a many-body system.
\end{abstract}


\maketitle

\section{Introduction}
\label{intro}

The description of quantum many-body systems, whether of nuclear, atomic or molecular nature,
is an everlasting challenge of theoretical physics \cite{Ring84,Fett1971}.
Even if the Hamiltonian is well known, the formalism needed to describe such systems can be baffling. 
One of the issues is that the actual inter-particle interactions in the medium can be very different from those in free space.
For strongly correlated many-body systems, ordinary perturbation
theory must be replaced by methods which perform an all-order summation of Feynman diagrams.
The self-consistent Green's functions (SCGF) formalism has been precisely devised 
to treat the correlated behavior of such systems \cite{Dick05}. 
The  time-dependent many-body field correlation functions, also called Green's functions (GFs) or propagators,
contain information associated with the addition or removal of a given number of particles
from the correlated ground state. Consequently, they can be used to obtain invaluable microscopic information 
on the many-particle system. More importantly, knowledge of the $N$-body GFs translates into the ability of computing all $N$-body operators and hence provides access to a wide range of observables. 
At the one-body level, the nonperturbative nature of the system is taken into account through the self-consistent solution 
of the Dyson equation~\cite{Mat92}.

The original many-body Green's functions formalism dates back to the 1960s~\cite{Mar59,Fett1971, Abr75}. 
In the past few decades, computational techniques have gradually improved to the point of allowing for 
fully \emph{ab-initio} studies that take into account beyond-mean-field correlations. 
First principles calculations are now routinely performed in solid state~\cite{Arya1998,Onida2002}, 
atomic and molecular physics \cite{Niessen1984,Ortiz2013,Barb07,Degr11} 
and nuclear structure~\cite{Muther2000,Dick04}.
In nuclear physics, finite nuclei have been studied using a variety of techniques, including the SCGF approach.
The Faddeev random phase approximation (FRPA) has been used to describe closed-shell isotopes \cite{Barb01,Barb09}. 
Medium-mass nuclei with open shells have been tackled within the Gorkov-Green's function method \cite{Soma2013}. 
The behavior of  correlations in infinite nuclear matter has been extensively studied using ladder summations~\cite{Ramos1989,Frick2003,Rios2008,Soma08}.

Initially, the many-body Green's functions framework was developed with Hamiltonians containing up to 
two-body (2B) interactions in mind. In the specific case of nuclear systems, however, three-body (3B) 
interactions play a substantial role. In infinite matter, three-body forces (3BFs) are long thought to be responsible for 
saturation 
\cite{grange1989,Zuo2002,Li2006,Soma08,Heb11,Lov12,Car13}. Ab-initio calculations of light nuclei have pointed towards the essential
role played by 3B interactions, particularly in reproducing the correct ground and excited-state properties \cite{Pieper2001,Wiringa2002,Navratil2007,Hagen07}. 
Recent breakthroughs in a wide variety of many-body techniques also 
indicate that 3B interactions play a role in medium-mass nuclei 
\cite{Ots10,Roth12,Hagen2012,Cipol13}. In spite of all these advances,  many-body theory with 
underlying 3B interactions has only been pushed forward in an intermittent fashion 
\cite{Morawetz2000,Hagen07,Heb10,Binder2013}. 

Our aim here is to develop the formal tools needed to include 3B interactions in nonperturbative calculations 
within the SCGF formalism. While our main motivation are nuclear systems, the formalism can be easily
applied to other many-body systems. 
Such an approach is pivotal both to provide theoretical foundations to approximations made so far and to 
advance the many-body formalisms for much-needed ab-initio nuclear structure. 
In the present paper, we present the extension of the SCGF formalism to include 3BFs by working out in full 
the first orders of the perturbation expansion and the self-consistent equations of motion.
Our approach is mainly practical. We want to put forward the basic rules that are needed to extend present
calculations to include 3B interactions. These include extensions in  the perturbative expansion, Feynman diagram
rules and the equation of motion (EOM) method. Moreover, we extend the Galitskii-Migdal-Koltun (GMK) sum-rule to 
compute the total energy of the system including such interactions \cite{Gal58,Kol74}. In principle, the approach is 
also able to incorporate 3B correlations in the many-body wave-function, but we will not discuss these explicitly here \cite{Rajaraman1967}. We will also not comment on the actual numerical implementation of the approach, 
which can be found elsewhere in the literature \cite{Cipol13,Car13}.

This study is made all the more timely in view of the notable recent efforts in improving the 
description of the strong interaction acting between nucleons. 
Realistic, phase-shift-equivalent 2B potentials have been traditionally used both in finite and infinite matter 
calculations 
\cite{Sto94,Wir95,Mac01}. Using such interactions as a starting point, however, one needs to evaluate the associated
3B forces consistently. Traditionally, this has been done either phenomenologically \cite{Piep01} or through an 
extension of the 
meson-exchange picture \cite{grange1989}. In the last decade, however, 
a more systematic approach has been devised, based 
on applying effective field theory to low-energy QCD \cite{Epel09,Mac11}. A particularly appealing
advantage of this approach is that it naturally gives rise to 2B, 3B and many-body interactions as the order of
the expansion increases. This avoids the somewhat \emph{ad hoc} adjustment of different ingredients in 2B and 3B 
potentials. 

Another important motivation to develop a formalism that explicitly includes 3BFs comes from recent
advances in developing low-momentum interactions \cite{Bog10}. These approaches reduce the computational effort
on the many-body side by taming the strong force using renormalization group techniques. However, this
results into the appearance of induced 3BFs (and other many-body forces). 
A number of approximations have been proposed to include both contributions in different many-body calculations~\cite{Hagen07,Heb10,Ots10,Heb11,Roth12,Cipol13,Car13}.
Most of these approaches involve, at some level, a normal ordering of the Hamiltonian to average on the third,
spectator particle. We will show that such approach is justified within the GFs formalism, provided that a class of 
interaction-irreducible diagrams are discarded to avoid double countings.
Incidentally, the latter approach goes beyond the usual normal ordering of perturbation theory
by incorporating fully correlated density matrices in the averaging procedure. 

Induced 3BFs are not exclusive to nuclear physics, as they arise from any truncation in the many-body model space. 
They play a role in a variety of other fields of physics \cite{Johnson2009,kievsky2011}. The 
bare and the induced 3B interactions, however, are treated on the same footing from a many-body perspective.
Hence, techniques that deal with 3BFs and 3B correlations within a many-body system are needed to describe such 
systems. The developments  proposed here are of relevance for applications on systems where induced many-body forces play a role. 

This paper is organized as follows. In Section \ref{section2} we exploit effective one-body (1B) and 2B potentials 
to group diagrams in the self-energy in a more compact form. We benchmark this approach by explicitly presenting 
the expansion of the single-particle (SP) GF up to third order in perturbation theory. 
We study, in Section \ref{section3}, the hierarchy of EOMs including 3BFs by means
of the interacting vertex $\Gamma$ functions. There, the
truncation of $\Gamma$ is presented up to second order in the perturbative expansion.
We also show that this truncation leads to the irreducible self-energy diagrams obtained in Section \ref{section2}.
Of relevance for practical applications, we present an 
improved ladder and ring summation that includes 3BFs. 
Section \ref{section4} is devoted to the GMK sum rule and its use for the calculation of the  ground state energy of the 
many-body system. Two extensions of this rule are presented which account for 3BFs. 
Appendix \ref{app_Feyn} provides a revision of Feynman diagrams and rules with 3BFs. We pay particular attention 
to additional symmetry factors due to equivalent groups of lines.
The proof for the expressions of the effective 1B and 2B interactions at arbitrary orders in perturbation theory
is presented in Appendix~\ref{app_Heff}. 

\section{Perturbative expansion of the one-body Green's functions}
\label{section2}

We work with a system of $N$ non-relativistic fermions interacting by means of 2B and 
3B interactions.
We divide the Hamiltonian into two parts, $\h H = \h H_0 + \h H_1$. 
$\h H_0 = \h T + \h U$ is an unperturbed, one-body contribution. It is 
the sum of the kinetic term and an auxiliary one-body operator, $\h U$, which defines the reference 
state for the perturbative expansion, $\rphizero$, on top of which correlations will be added
\footnote{A typical choice in nuclear physics would be a Slater determinant of single-particle harmonic oscillator or a Woods-Saxon wave function.}.  
The second term of the Hamiltonian, 
$\h H_1 = -\h U + \h V + \h W$, includes the interactions. $\h V$ and $\h W$ denote, respectively, the two- and three-body interaction operators.
In a second-quantized framework, the full Hamiltonian reads:
\beqn
\label{H}
\h H &=& \sum_{\al} \vep^0_\al\, a^\dg_\al a_\al - \sum_{\al\be}U_{\al\be}\, a^\dg_\al a_{\be}
\\\nn &+&
\f 1 4 \sum_{\substack{\al\ga\\\be\de}}V_{\al\ga,\be\de}\, a_\al^\dg a_\ga^\dg a_{\de} a_{\be}
\\\nn &+& 
\frac{1}{36}\sum_{\substack{\al\ga\ep \\ \be\de\eta}} W_{\al\ga\ep,\be\de\eta}\,
a_\al^\dg a_\ga^\dg a_\ep^\dg a_{\eta} a_{\de} a_{\be} \, .
\enqn
The greek indices $\al$,$\be$,$\ga$,\ldots label a complete set of SP states which diagonalize the unperturbed Hamiltonian, $\h H_0$, with eigenvalues $\vep_\al^0$. 
$a^\dg_\al$ and $a_\al$ are creation and annihilation operators for a particle in state $\al$. 
The matrix elements of the 1B operator $\h U$ are given by $U_{\al\be}$. Equivalently, the matrix elements of the
2B and 3B forces are $V_{\al\ga,\be\de}$ and $W_{\al\ga\ep,\be\de\eta}$. 
In the following, we work with antisymmetrized matrix elements in both the 2B and the 3B sectors.

The main ingredient of our formalism is the 1B GF, also called SP propagator or  2-point GF, 
which provides a complete description of one-particle and one-hole excitations of the many-body system. 
More specifically, the SP propagator is defined as the expectation value 
of the time-ordered product of an annihilation and a creation operators
in the Heisenberg picture:
\beq
\label{G}
\ii\hbar\, G_{\al\be} (t_\al-t_\be)
=\langle\Psi_0^N|{\cal T}[a_\al(t_\al)a_{\be}^\dg(t_\be)]|\Psi_0^N\rangle \,,
\enq
where $|\Psi_0^N\rangle$ is the interacting $N$-body ground state of the system. 
The time ordering operator brings operators with earlier times to the right, with the corresponding fermionic
permutation sign.
For $t_\al-t_\be>0$, this results in the addition of a particle to the state $\be$ at time $t_\be$ 
and its removal from state $\al$ at time $t_\al$. Alternatively, for $t_\be-t_\al>0$,
the removal of a particle from state $\al$ occurs at time $t_\al$ and its addition to state $\be$ at time $t_\be$.
These correspond, respectively, to the propagation of a particle or a hole excitation through the system.
We can also introduce the 4-point and 6-point GFs:
\beqn
\label{g4pt}
&&
\ii\hbar \,G^{4-{\rm pt}}_{\al\ga,\be\de}(t_\al,t_\ga;t_{\be},t_{\de}) = 
\\\nn &&\quad\quad
\langle\Psi_0^N|{\cal T}[a_\ga(t_\ga)a_\al(t_\al)
a_{\be}^\dg(t_{\be})a_{\de}^\dg(t_{\de})]|\Psi_0^N\rangle \,,
\enqn
\beqn
\label{g6pt}
&&
\ii\hbar \,G^{6-{\rm pt}}_{\al\ga\ep,\be\de\eta}(t_\al,t_\ga,t_\ep;t_{\be},t_{\de},t_{\eta}) = 
\\\nn &&\quad
\langle\Psi_0^N|{\cal T}[a_\ep(t_\ep)a_\ga(t_\ga)a_\al(t_\al)
a_{\be}^\dg(t_{\be})a_{\de}^\dg(t_{\de})a_{\eta}^\dg(t_{\eta})]|\Psi_0^N\rangle \, .
\enqn
Physically, the interpretation of Eq.~(\ref{g4pt}) and Eq.~(\ref{g6pt}) follows that of the 2-point GF in Eq.~(\ref{G}). 
In these cases, more combinations of particle and hole excitations are encountered depending on the ordering
of the several time arguments.
Note also that these propagators provide access to all 2B and 3B observables.
The extension to formal expressions for higher many-body GFs is straightforward.

In the following, we will consider propagators both in time representation, as defined above, or
in energy representation. 
Note that, due to time-translation invariance, the \hbox{$m$-point} GF depends only on
$m-1$ time differences or, equivalently, $m-1$ independent frequencies. Hence the Fourier transform to
the energy representation is only well-defined when the total energy is conserved:
\begin{widetext}
\beqn
\label{Gmpt_ft}
&& 2\pi \delta (\omega_\al + \omega_\ga + \ldots  - \omega_\be - \omega_\de - \ldots) \times 
G^{m-{\rm pt}}_{\al \ga \ldots,  \be \de  \ldots} (\omega_\al, \omega_\ga, \ldots ; \omega_\be , \omega_\de ,  \ldots ) = 
 \\
&& ~~ \int \hspace{-1mm} \d t_\al   \int \hspace{-1mm} \d t_\ga \; \ldots 
                \int \hspace{-1mm} \d t_\be  \int \hspace{-1mm} \d t_\de \ldots
                \;   e^{i(\omega_\al t_\al + \omega_\ga t_\ga + \ldots)}
G^{m-{\rm pt}}_{\al \ga \ldots,  \be \de  \ldots} (t_\al, t_\ga, \ldots; t_\be , t_\de ,  \ldots ) \,
e^{-i(\omega_\be t_\be + \omega_\de t_\de + \ldots)} \; .
\nn
\enqn
\end{widetext}
For the 1B GFs, one also considers,
\beq
\label{G1B_ft}
G_{\al \be} (\omega) = \int d\t \; e^{i \omega \t} G_{\al \be}(\t) = G^{2-pt} _{\al \be} (\omega, \omega) \; .
\enq

Interactions in the many-body system can be treated by means of a perturbative expansion. 
For the 1B propagator, $G$, this expansion reads \cite{Mat92,Dick05}:
\begin{widetext}
\beqn
\label{gpert}
G_{\al\be}(t_\al-t_\be) = -\f \ii \hbar \sum_{n=0}^\infty \left(-\f \ii\hbar\right)^n\frac{1}{n!}\int \hspace{-1mm} \d t_1 \; \ldots \int \hspace{-1mm} \d t_n \lphizero\T[\h H_1(t_1) \ldots \h H_1(t_n)a^I_\al(t_\al){a_{\be}^I}^\dg(t_\be)]\rphizero_\text{conn} \; , &&
\enqn
\end{widetext}
where $\rphizero$ is the unperturbed many-body ground state, i.e. our reference $N$-body state. 
$a^I_\al$ and ${a_{\be}^I}^\dg$ are now operators in the interaction picture with respect to $H_0$. 
The subscript ``conn'' implies that only \emph{connected} diagrams have to be considered when performing
the Wick contractions of the time-ordered product.

$H_1$ contains contributions from 1B, 2B and 3B interactions. Thus, the expansion involves terms
with individual contributions of each force, as well as combinations of these. 
Feynman diagrams are essential to keep track of such a variety of different contributions.
The set of Feynman diagrammatic rules that stems out of Eq.~(\ref{gpert}) in the presence of 3B interactions
is reviewed in detail in Appendix~\ref{app_Feyn}. In general, these are unchanged with respect to the 2B case. 
However, we provide a few examples to illustrate the appearance of non-trivial symmetry factors when 3B are 
considered. This complicates the rules of the symmetry factors and illustrates some of the difficulties associated with 
many-particle interactions. In the following, we will work mostly with unlabelled Feynman diagrams. 
We also work with antisymmetrized matrix elements but, in contrast to Hugenholtz diagrams \cite{Blaizot86},
we expand the interaction vertices and show them with different types of lines for clarity. 

A first reorganization of the contributions generated by Eq.~(\ref{gpert}) is obtained by considering 
\emph{one-particle reducible} diagrams, i.e. diagrams that can be disconnected by cutting a single fermionic line. 
Reducible diagrams are generated by an all-orders summation through Dyson's equation~\cite{Dick05},
\beqn
G_{\al\be}(\om)=G^{(0)}_{\al\be}(\om)+\sum_{\ga \de}G^{(0)}_{\al\ga}(\om)\Sigma^\star_{\ga\de}(\om)G_{\de\be}(\om)\, .
\label{Dyson}
\enqn
Thus, in practice, one only needs to include \emph{one-particle irreducible} (1PI) contributions
to the self-energy, $\Sigma^\star$. 
The uncorrelated SP propagator, $G^{(0)}$, is associated with the system governed by the $H_0$ Hamiltonian 
and represents the $n=0$ order in the expansion of Eq.~(\ref{gpert}). 
In the previous equation, $\omega$ corresponds to the energy of the propagating particle or hole excitation.
The irreducible self-energy $\Sigma^\star$, appearing in Eq.~(\ref{Dyson}), describes the kernel of all 1PI diagrams.
This operator plays a central role in the GF formalism and can be interpreted as the 
non-local and energy-dependent interaction that each fermion feels due to the interaction with the medium. 
At positive energies, $\Sigma^\star(\omega)$ is also identified with the optical potential for scattering of a particle 
from the many-body target~\cite{Blaizot86,Capuzzi1996,Cederbaum2001,Bar05,Cha06}. 

A further level of simplification in the self-energy expansion 
can be obtained if unperturbed propagators, $G^{(0)}$, in the internal fermionic lines are replaced by dressed GFs, $G$. 
This process is generically called propagator renormalization and 
further restricts the set of diagrams to \emph{skeleton} diagrams \cite{Blaizot86,Dick05}.
These are defined as 1PI diagrams that do not contain 
any portion that can be disconnected by cutting a fermion line twice at any two different points. 
These portions would correspond to self-energy insertions, which are already re-summed into the dressed propagator $G$ by Eq.~(\ref{Dyson}).
The SCGF approach is precisely based on diagrammatic expansions of such skeleton diagrams 
with renormalized propagators.

In principle, this framework offers great advantages. First, it is intrinsically nonperturbative and completely
independent from any choice of the reference state and auxiliary 1B potential, $\hat{U}$, 
which automatically drops out of Eq.~(\ref{Dyson}). 
Second, many-body correlations are expanded directly in terms of SP excitations which are closer
to the exact solution than those associated with the unperturbed state, $\rphizero$. Third, 
the number of diagrams is vastly reduced to 1PI skeletons. 
Fourth, a full SCGF calculation automatically satisfies the basic conservation
laws~\cite{Baym:1961zz,Baym:1962sx,Dick05}. 
In practice, however, calculating diagrams with dressed propagators is computationally
more expensive than using the plain $G^{(0)}$  in perturbation theory. 
Moreover, self-consistency requires an iterative solution
for $\Sigma^\star$ and for $G$ via the Dyson equation, Eq.~(\ref{Dyson}). 
Therefore, the SCGF scheme is not always applied in full detail, but it is often employed to provide
important guidance in developing working approximations to the self-energy.

\subsection{Interaction-irreducible diagrams}
 \label{sec_irr_diags}

\begin{figure}
\begin{center}
\includegraphics[width=1\columnwidth]{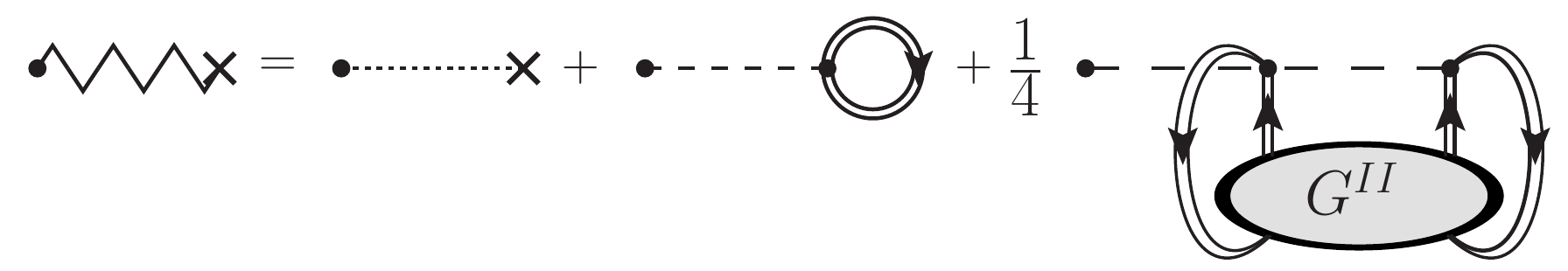}
\caption{Diagrammatic representation of the effective 1B interaction of Eq.~(\ref{ueff}). This is given by the sum of the original 1B potential (dotted line), the 2B interaction (dashed line) contracted with a dressed SP propagator, $G$ (double line with arrow), and the 3B interaction (long-dashed line) contracted with a dressed 2B propagator $G^{II}$. The correct symmetry factor of 1/4 in the last term is also shown explicitly. }
\label{ueffective}
\end{center}
\end{figure}

\begin{figure}
\begin{center}
\includegraphics[width=0.90\columnwidth]{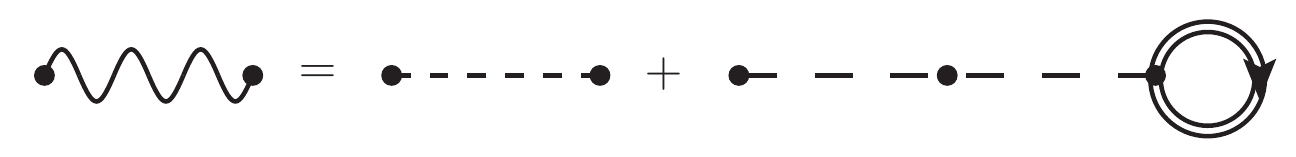}
\caption{Diagrammatic representation of the effective 2B interaction of Eq.~(\ref{veff}). This is given by the sum 
of the original 2B interaction (dashed line) and the
3B interaction (long-dashed line) contracted with a dressed SP propagator,  $G$.  }
\label{veffective}
\end{center}
\end{figure}

It is possible to further restrict the set of relevant diagrams by
exploiting the concept of \emph{effective interactions}. 
Let us consider an articulation vertex in a generic Feynman diagram. 
A 2B,  3B or higher interaction vertex is an articulation vertex if, when cut, it gives rise to two disconnected diagrams \footnote{1B vertices cannot be split and therefore cannot be articulations.}
Formally, a diagram is said to be \emph{interaction-irreducible} if it contains no articulation vertices.
Equivalently, a diagram is interaction reducible 
if there exist a group of fermion lines (either interacting or not) that leave one
interaction vertex and eventually all return to it.

When an articulation vertex is cut, one is left with a cycle of fermion lines that all connect to
the same interaction. If there were $p$ lines connected to this interaction vertex, this set 
of closed lines would necessarily be part of a $2p$-point GF
\footnote{
More specifically, these fermion lines contain an instantaneous contribution of the many-body 
GF that enters and exits the same interaction vertex, corresponding to a $p-$body reduced density matrix.}.
If this GF is computed explicitly in the calculation,
one can use it to evaluate all these contributions straight away. This eliminates the need
for computing all the diagrams looping in and out of the articulation vertex, at the 
expense of having to find the many-body propagator. 
An $n$-body interaction vertex with $p$ fermion lines looping over it is an $n-p$ 
effective interaction operator.
Infinite sets of interaction-reducible diagrams can be sub-summed by means of 
effective interactions. 

The two cases
of interest when 2B and 3B forces are present in the Hamiltonian are shown in Figs.~\ref{ueffective} and~\ref{veffective}
that give, respectively, the diagrammatic definition of  the 1B and 2B effective interactions. 
The 1B effective interaction is obtained by adding up three contributions: the original 1B interaction; 
a 1B average over the 2B interaction; 
and a 2B average over the 3B force. The 1B and 2B averages are performed using fully dressed 
propagators.
Similarly, an effective 2B force is obtained from the original 2B interaction plus a 1B average over the
3B force. Note that these go beyond usual normal-ordering ``averages" in that they are performed
over fully-correlated, many-body propagators. Similar definitions would hold for higher-order forces and
effective interactions beyond the 3B level. 

Hence, for a system with up to 3BFs, we define an effective Hamiltonian,
\beq
\widetilde H_1= {\widetilde U} + {\widetilde  V} + \h W \,
\label{Heff}
\enq
where $\widetilde U$ and  $\widetilde V$ represent effective interaction operators. 
The diagrammatic expansion arising from  Eq.~(\ref{gpert}) with the effective Hamiltonian $\widetilde H_1$ is
formed only of (1PI, skeleton) interaction-irreducible diagrams to avoid any possible double counting.
Note that the 3B interaction, $\h W$, remains the same as in Eq.~(\ref{H}) but enters only the 
interaction-irreducible diagrams with respect to 3B interactions.
The explicit expressions for the 1B and 2B effective interaction operators are:
\begin{widetext}
\beqn
\label{ueff}
\widetilde U& =&\sum_{\al\be}\Big[- U_{\al\be} 
- \ii\hbar \sum_{\ga\de}V_{\al\ga,\be\de} \, G_{\de \ga}(t-t^+) 
+ \f{\ii\hbar}{4} \sum_{\substack{\ga\ep \\ \de\eta}} W_{\al\ga\ep,\be\de\eta}
\,G^{II}_{\de\eta , \ga\ep}(t-t^+)\Big] a_\al^\dg a_{\be}\,, \\ 
\label{veff}
\widetilde V &=& \f 1 4\sum_{\substack{\al\ga\\\be\de}}\left[V_{\al\ga,\be\de}
- \ii\hbar \sum_{\ep\eta}W_{\al\ga\ep,\be\de\eta} \,G_{\eta\ep}(t-t^+)\right] a_\al^\dg a_\ga^\dg a_{\de}a_{\be} \, .
\enqn
\end{widetext}
We have introduced a specific component of the 4-point GFs,
\beq
G^{II}_{\de\eta , \ga\ep}(t-t') = G^{4-\rm{pt}}_{\de\eta , \ga\ep}(t^+, t; t', t'^+) \, ,
\label{g2g4pt}
\enq
which involves two-particle and two-hole propagation. 
This is the so-called two-particle and two-time
Green's function. 
Let us also note that the contracted propagators in Eqs.~(\ref{ueff}) and (\ref{veff}) correspond to 
the full 1B and 2B reduced density matrices of the many-body system:
\beqn
\label{1B_densitymatrix}
\rho^{1B}_{\de\ga} &= ~ \langle\Psi_0^N| \, a_{\ga}^\dg a_{\de} \, |\Psi_0^N\rangle \quad &=  -\ii\hbar\, G_{\de\ga}(t-t^+) \; ,
\\
\rho^{2B}_{\de\eta , \ga\ep} &= \langle\Psi_0^N| \, a_{\ga}^\dg a_{\ep}^\dg a_{\eta} a_{\de} \, |\Psi_0^N\rangle \; &=
~\ii\hbar\,G^{II}_{\de\eta , \ga\ep}(t-t^+) \, . \qquad
\enqn
In a self-consistent calculation, effective interactions should be computed iteratively at each step, 
using correlated 1B and 2B propagators as input.

The effective Hamiltonian of Eq.~(\ref{Heff})  not only regroups Feynman diagrams in 
a more efficient way, but also defines the effective 1B and 2B terms from 
higher order interactions. Averaging the 3BF over one and two spectator particles
in the medium is expected to yield the most important contributions to the
many-body dynamics in nuclei~\cite{Hagen07,Roth12}.
We note that Eqs.~(\ref{ueff}) and~(\ref{veff}) are exact and can be derived rigorously 
from the perturbative expansion. Details of the proof are discussed in Appendix~\ref{app_Heff}. 
As long as interaction-irreducible diagrams are used together with the 
effective Hamiltonian, $\widetilde{H}_1$, this approach provides a systematic
way to incorporate many-body forces in the calculations and to 
generate effective in-medium interactions. More importantly, the formalism is such that 
symmetry factors are properly considered and no diagram is over-counted.

This approach can be seen as a generalization of the normal
ordering of the Hamiltonian with respect to the reference state $\rphizero$, 
a procedure that is already been
used in nuclear physics applications with 3BFs~\cite{Hagen07,Bog10,Roth12}.
In both the traditional normal ordering and our approach, 
the ${\widetilde{U}}$ and ${\widetilde{V}}$ operators contain
contributions from higher order forces, while $\hat{W}$ remains unchanged. The
normal ordered interactions affect only excited configurations
with respect to $\rphizero$, but not the reference state itself. Similarly, the effective
operators discussed above only enter interaction-irreducible diagrams.
As a matter of fact, if the unperturbed 1B and 2B propagators were 
used in Eqs.~(\ref{ueff}) and~(\ref{veff}),
the effective operators ${\widetilde{U}}$ and ${\widetilde{V}}$ would
trivially reduce to the contracted 1B and 2B terms of normal ordering.
In the present case, however, the contraction goes beyond normal ordering
because it is performed with respect to the exact
correlated density matrices. To some extent, one can think of 
the effective Hamiltonian, $\widetilde{H}$,  as being reordered with respect
to the interacting many-body ground-state $|\Psi_0^N\rangle$, rather than 
the non-interacting  $\rphizero$. 
This effectively incorporates correlations that, in the normal ordering procedure,
must be instead calculated explicitly by the many-body approach.
Calculations indicate that such correlated averages are important in both the saturation
mechanism of nuclei and nuclear matter \cite{Cipol13,Car13}.

Note that a normal ordered Hamiltonian also contains a 0B term equal to
the expectation value of the original Hamiltonian, $\hat{H}$, with respect to $\rphizero$.
Likewise, the full contraction of $\hat{H}$, according to the procedure of Appendix~\ref{app_Heff},
will yield the exact ground state energy:
\beqn
\nn
E_0^N  &=&-\ii\hbar \,\sum_{\al\be} T_{\al\be} \, G_{\be\al}(t-t^+)
\\\nn &&+
\f{\ii\hbar}{4}\sum_{\substack{\al\ga\\\be\de}}V_{\al\ga,\be\de}\, G^{II}_{\be\de,\al\ga}(t-t^+) 
\\\nn && - \f{\ii\hbar}{36}
\sum_{\substack{\al\ga\ep \\ \be\de\eta}} W_{\al\ga\ep , \be\de\eta}  \, G^{III}_{\be\de\eta,\al\ga\ep}(t-t^+)
\\ &=& \langle\Psi_0^N| \, H  \, |\Psi_0^N\rangle \,,
\label{e0}
\enqn
in accordance with our analogy between the effective Hamiltonian, $\widetilde{H}=\h H_0+\widetilde{H}_1$,
and the usual normal ordering.

Before we move on, let us mention a subtlety arising in the Hartree-Fock (or lowest-order) approximation to the 
two-body propagator.  If one were to insert $\widetilde{V}$ into the second term of the right hand side of 
Eq.~(\ref{ueff}), one would introduce a double counting of the pure 3BF Hartree-Fock component. 
This is forbidden because the diagram in question would be interaction reducible. 
The correct  3BF Hartree-Fock term is actually included as part of the last term of  Eq.~(\ref{ueff}) 
(see also Fig.~\ref{ueffective}). Consequently, there is no Hartree-Fock term arising from the effective
interactions. Instead, this lowest-order contribution is fully taken into account within the 1B effective
interaction.

\subsection{Self-energy expansion up to third order}
 \label{sec_PT_3ord}
 
As a demonstration of the simplification introduced by the effective interaction approach,
in this subsection we will derive all interaction-irreducible
contributions to the proper self-energy up to third order in perturbation theory. 
We will discuss these contributions order by order, thus providing an overview of how the approach can be extended
to higher-order perturbative and also to nonperturbative calculations.
Among other things,
this exercise will illustrate the amount and variety of new diagrams that need to be 
considered when 3BFs are used.

For a pure 2B Hamiltonian, the only possible interaction-reducible contribution to the self-energy 
is the generalized Hartree-Fock diagram. This corresponds to the second term on the right hand 
side of Eq.~(\ref{ueff}) (see also Fig.~\ref{ueffective}). Note that this can go beyond the usual 
Hartree-Fock term in that the internal propagator is dressed. 
This diagram appears at first order in any SCGF expansion and it is
routinely included in most GF calculations with 2B forces. Thus, regrouping diagrams 
in terms of effective interactions, such as Eqs.~(\ref{ueff}) and~(\ref{veff}), 
give no practical advantages unless 3BFs (or higher-body forces) are present. 

If 3BFs are considered, the only first-order, interaction-irreducible contribution 
is precisely given by the one-body effective interaction depicted in Fig.~\ref{ueffective},
\beq
\Sigma^{\star , (1)}_{\al \be} = \widetilde{U}_{\al\be} \; . 
\label{eq:1ord}
\enq
Since $\widetilde{U}$ is in itself a self-energy insertion, it will not appear in any other, higher-order skeleton
diagram. Even though it only contributes to Eq.~(\ref{eq:1ord}), the
effective 1B potential is very important since it determines the energy-independent part
of the self-energy. It therefore represents the (static) mean-field seen by every particle, due to both 
2B and 3B interactions.
As already mentioned, Eq.~(\ref{ueff}) shows that this potential incorporates three separate terms, including
the Hartree-Fock potentials due  to both 2B and 3BFs and higher-order, interaction-reducible 
contributions due to the dressed $G$ and $G^{II}$ propagators. 
Thus, even the calculation of this lowest-order
term $\Sigma^{\star , (1)}$ requires an iterative procedure to evaluate the internal many-body 
propagators self-consistently.

Note that, if one were to stop at the Hartree-Fock level, the 4-point GF would reduce to the direct and
exchange product of two 1B propagators. In that case, the last term of Eq.~(\ref{ueff}) (or Fig.~\ref{ueffective})
would reduce to the pure 3BF Hartree-Fock contribution with the correct $1/2$ factor in front, due to the two
equivalent fermionic lines. This approximate treatment of the 2B propagator in the 1B effective interaction
has been employed in most nuclear physics calculations up to date, including both finite nuclei
\cite{Ots10,Roth12,Cipol13} and nuclear matter \cite{Soma09,Heb10,Heb11,Lov12,Li12,Car13} applications.
Numerical implementations of  averages with fully correlated 2B propagators are underway.

\begin{figure}[t]
  \centering
  \subfloat[]{\label{2ord_2B}\includegraphics[scale=0.55]{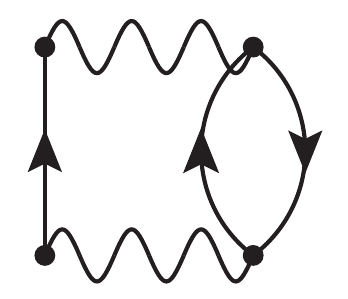}}
  \hspace{2cm}
  \subfloat[]{\label{2ord_3B}\includegraphics[scale=0.55]{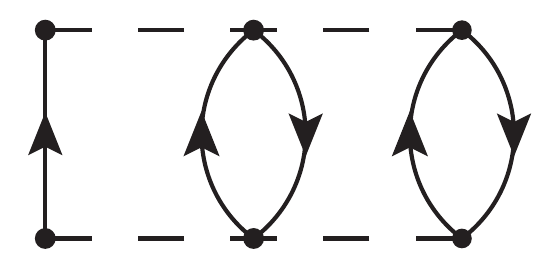}}
  \caption{1PI, \emph{skeleton} and \emph{interaction-irreducible} self-energy diagrams appearing at second order in the perturbative expansion of Eq.~(\ref{gpert}), using the effective Hamiltonian of Eq.~(\ref{Heff}).}
  \label{2ord}
\end{figure}

\begin{figure}
  \centering
  \hspace{.5cm}
  \subfloat[]{\label{2ord_2B_split_a}\includegraphics[scale=0.55]{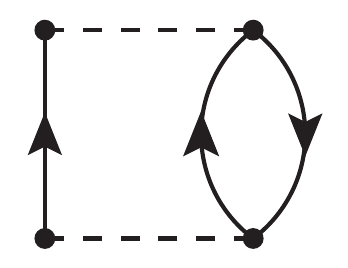}}
  \hspace{1.5cm}
  \subfloat[]{\label{2ord_2B_split_b}\includegraphics[scale=0.55]{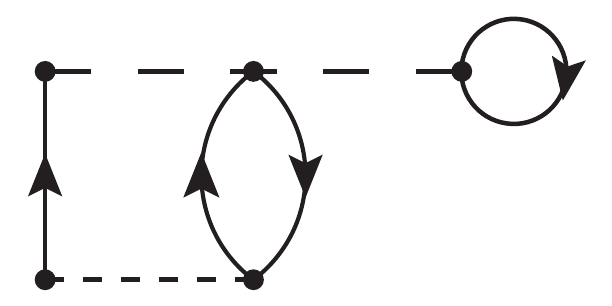}}
  \hspace{1cm}  \newline  \vskip .5cm
    \subfloat[]{\label{2ord_2B_split_c}\includegraphics[scale=0.55]{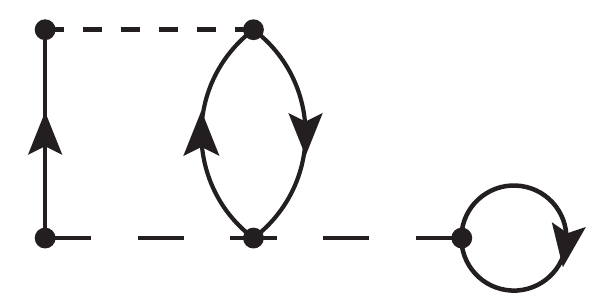}}
    \hspace{1cm} 
  \subfloat[]{\label{2ord_2B_split_d}\includegraphics[scale=0.55]{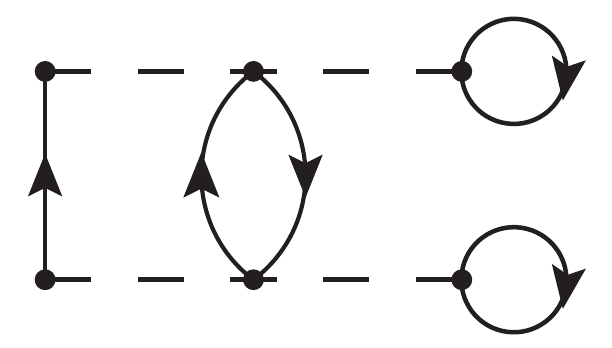}}
  \caption{These four diagrams are contained in diagram~\ref{2ord_2B}. 
 They correspond to one 2B \emph{interaction-irreducible} diagram, (a), and
 three \emph{interaction-reducible} diagrams, (b) to (d). }
\label{2ord_2B_split}
\end{figure}

At second order, there are only two interaction-irreducible diagrams, that we show
in Fig.~\ref{2ord}. Diagram \ref{2ord_2B} has the same structure as the well-known contribution due to
2BFs only, involving two-particle--one-hole ($2p1h$) and two-hole--one-particle ($2h1p$) intermediate states.
This diagram, however, is computed with the 2B effective interaction (notice the wiggly line) instead
of the original 2B force and hence it corresponds to further interaction-reducible diagrams. 
By expanding the effective 2B interaction according to Eq.~(\ref{veff}), the 
contribution of Fig.~\ref{2ord_2B} splits into the four diagrams of
 Fig.~\ref{2ord_2B_split}~[see also a similar example in Fig.~\ref{rule9-2a}].
 
 The second interaction-irreducible diagram arises from explicit 3BFs and it  is given in 
Fig.~\ref{2ord_3B}. 
One may expect this contribution to play a minor role due to phase space arguments,
as it involves  $3p2h$ and $3h2p$ excitations at higher excitation energies. 
Moreover,  3BFs are generally weaker than the
corresponding  2BFs (typically, $<\widehat{W}>\approx\f{1}{10}<\widehat{V}>$ for nuclear
interactions~\cite{grange1989,Epel09}). 
Summarizing, at second order in standard self-consistent perturbation theory, one would
find a total of 5 skeleton diagrams. Of these, only 2 are interaction irreducible 
and need to be calculated when effective interactions are considered.

\begin{figure*}
  \centering
  \subfloat[]{\label{3ord_2B_1}\includegraphics[scale=0.50]{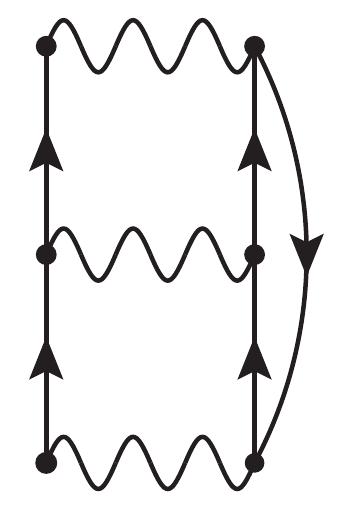}}
  \hspace{2cm}
  \subfloat[]{\label{3ord_2B_2}\includegraphics[scale=0.50]{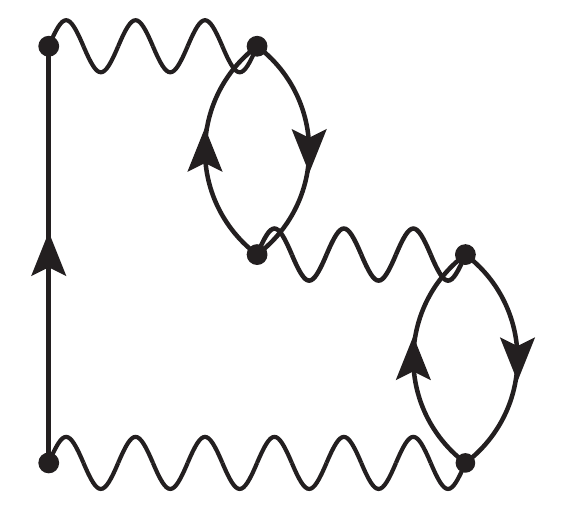}}
  \hspace{3cm}
  \subfloat[]{\label{3ord_232B}\includegraphics[scale=0.50]{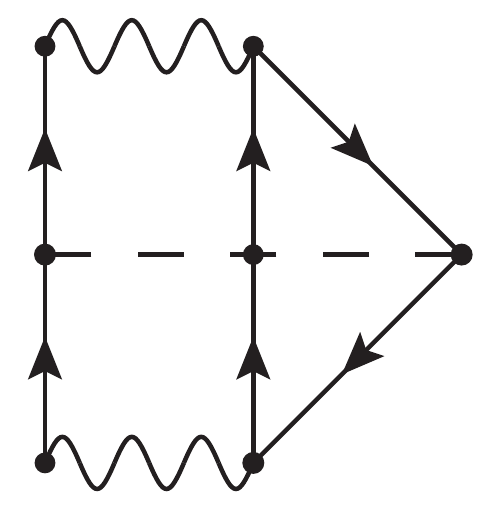}}
  \newline    \vskip .7cm
  \subfloat[]{\label{3ord_223B_1}\includegraphics[scale=0.50]{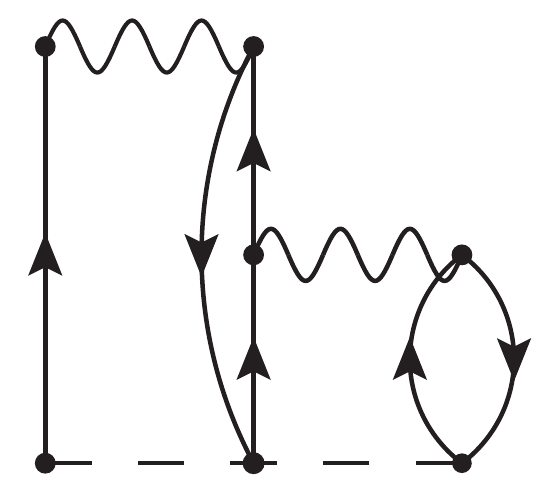}}
  \hspace{1.5cm}
  \subfloat[]{\label{3ord_223B_2}\includegraphics[scale=0.50]{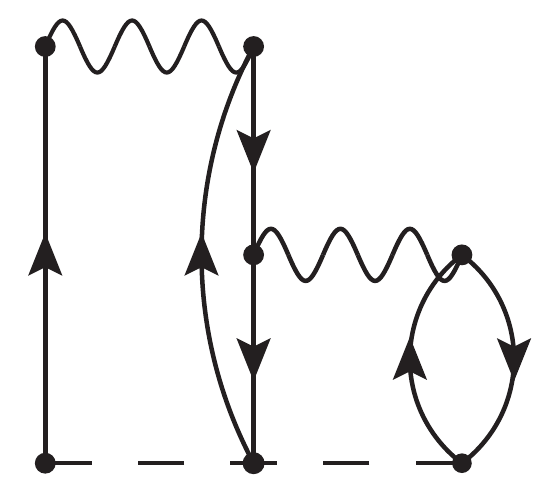}}
  \hspace{1.5cm}
   \subfloat[]{\label{3ord_322B_1}\includegraphics[scale=0.50]{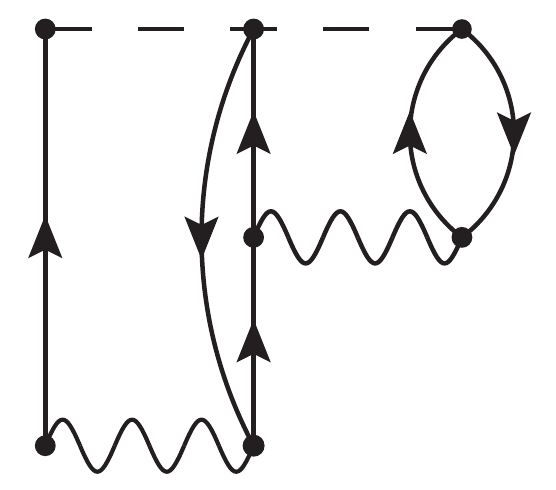}}
  \hspace{1.5cm}
  \subfloat[]{\label{3ord_322B_2}\includegraphics[scale=0.50]{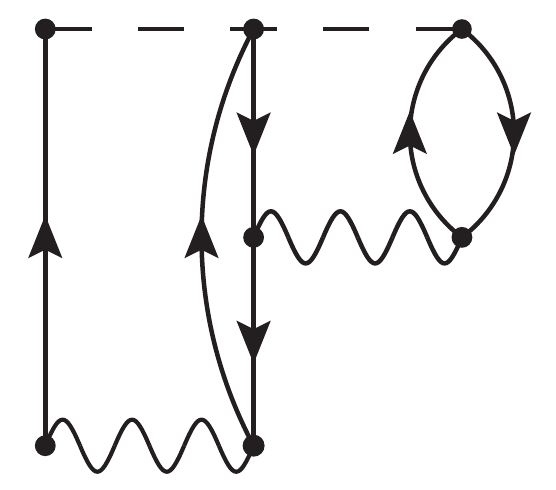}}
  \newline   \vskip .7cm
  \subfloat[]{\label{3ord_233B_1}\includegraphics[scale=0.50]{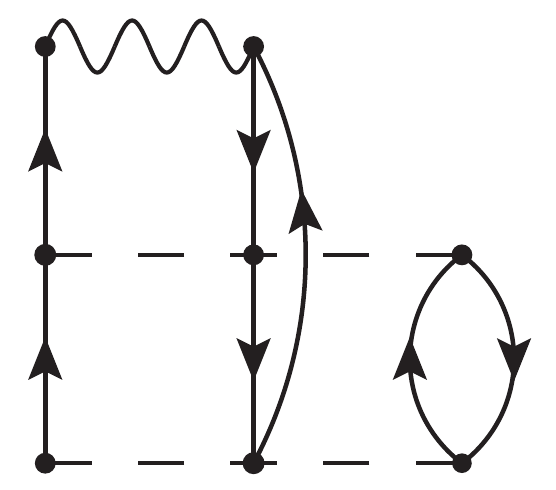}}
  \hspace{1.5cm}
  \subfloat[]{\label{3ord_233B_2}\includegraphics[scale=0.50]{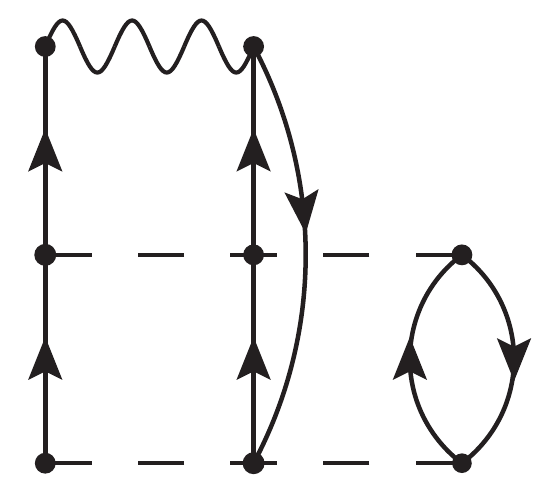}}
  \hspace{1.5cm}
  \subfloat[]{\label{3ord_332B_1}\includegraphics[scale=0.50]{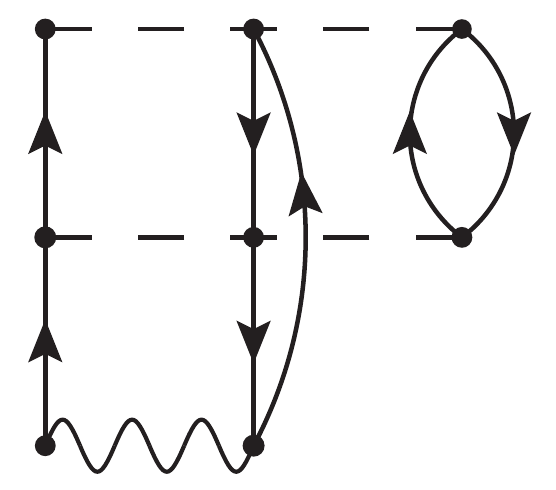}}
  \hspace{1.5cm}
  \subfloat[]{\label{3ord_332B_2}\includegraphics[scale=0.50]{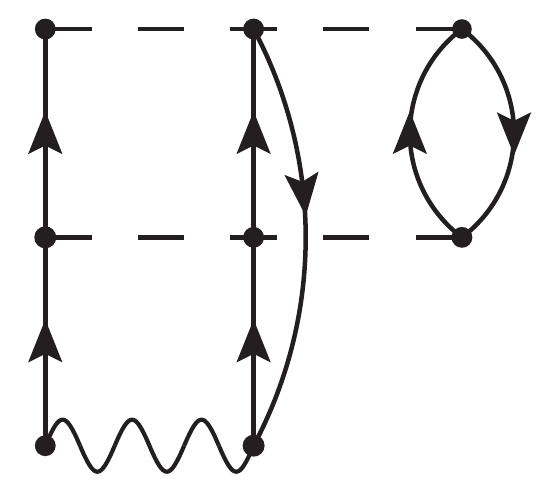}}
  \newline   \vskip .7cm
  \subfloat[]{\label{3ord_323B_1}\includegraphics[scale=0.50]{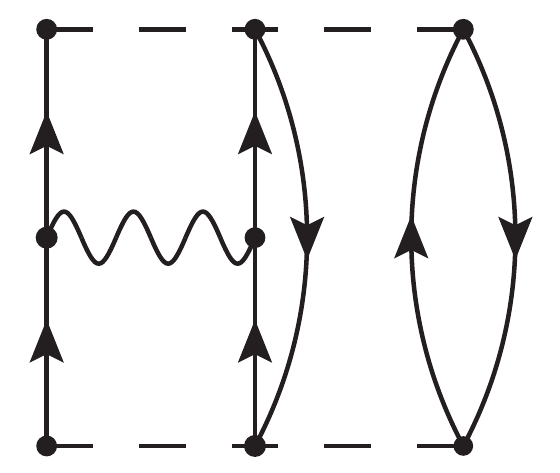}}
  \hspace{1.5cm}
  \subfloat[]{\label{3ord_323B_2}\includegraphics[scale=0.50]{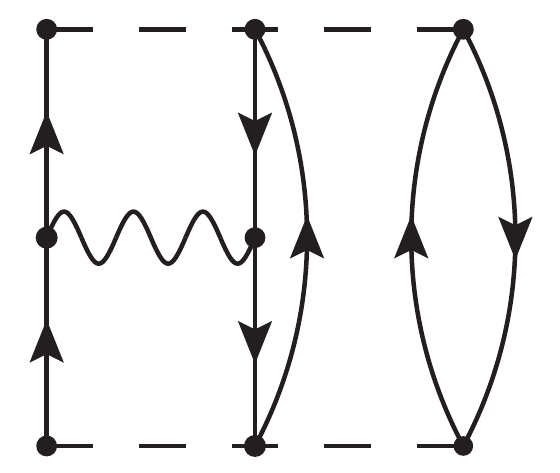}}
  \hspace{1.5cm}
  \subfloat[]{\label{3ord_323B_3}\includegraphics[scale=0.50]{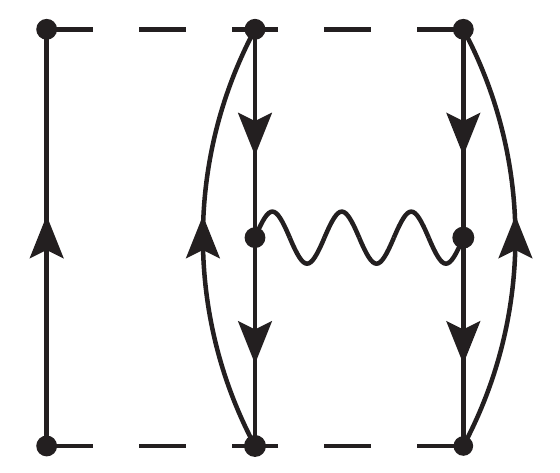}}
  \hspace{4cm}
  \newline   \vskip .7cm
  \hspace{2.5cm}
  \subfloat[]{\label{3ord_333B_1}\includegraphics[scale=0.55]{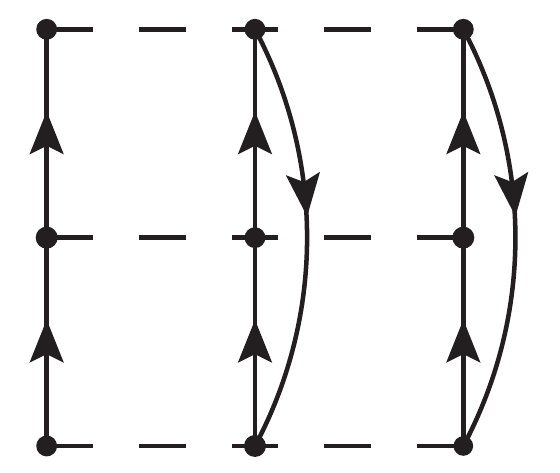}}
  \hspace{1.5cm}
  \subfloat[]{\label{3ord_333B_2}\includegraphics[scale=0.55]{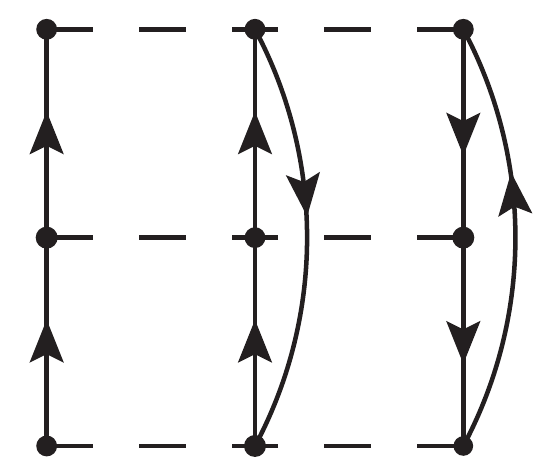}}
  \hspace{1.5cm}
  \subfloat[]{\label{3ord_333B_3}\includegraphics[scale=0.55]{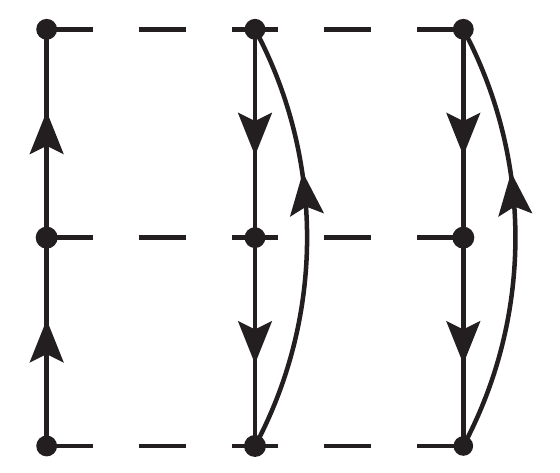}}
   \caption{1PI, \emph{skeleton} and \emph{interaction-irreducible} self-energy diagrams appearing at third order in the perturbative expansion of Eq.~(\ref{gpert}) using the effective Hamiltonian of Eq.~(\ref{Heff}).} 
  \label{3ord}
\end{figure*}

Fig.~\ref{3ord} shows all the 17 interaction-irreducible diagrams appearing at third order. 
Again, note that, expanding the effective interaction $\widetilde{V}$, would generate a much 
larger number of diagrams (53 in total). 
Diagrams~\ref{3ord_2B_1} and~\ref{3ord_2B_2} are the only third order terms that would appear in the 2BF case. 
Numerically, these two diagrams only require evaluating Eq.~(\ref{veff}) beforehand, but can otherwise be dealt with 
using existing 2BF codes. They have already been exploited to include 3BFs in nuclear structure studies~\cite{Soma08,Heb10,Heb11,Cipol13,Car13}. 

The remaining 15 diagrams, from~\ref{3ord_232B} to~\ref{3ord_333B_3}, appear when 3BFs are introduced. 
These third-order diagrams are ordered in  Fig.~\ref{3ord} in terms of increasing numbers of 3B interactions and, within these,
in terms of increasing number of particle-hole excitations. 
Qualitatively, one would expect that this should correspond to a decreasing importance of their contributions.
Diagrams~\ref{3ord_2B_1}-\ref{3ord_232B}, for instance, only involve $2p1h$ and $2h1p$ intermediate configurations, normally needed to describe particle addition and removal energies to dominant quasiparticle peaks as well as total ground state energies. 

Diagram~\ref{3ord_232B} includes one 3B irreducible interaction term and still needs to be 
investigated within the SCGF method. Normal-ordered Hamiltonian studies~\cite{Hagen07,Roth12} 
clearly suggest that this brings in a small correction to the total energy with respect to 
diagrams~\ref{3ord_2B_1} and~\ref{3ord_2B_2}. 
This is in line with the qualitative analysis of the number of $\widetilde{V}$ and $\h{W}$ 
interactions entering these diagrams.
Diagrams~\ref{3ord_2B_1}-\ref{3ord_232B} all represent the first order term in an all order summation 
needed to account for configuration mixing between $2p1h$ or $2h1p$ excitations. 
Nowadays, resummations of these configurations are performed routinely for the first two 
diagrams in third-order algebraic diagrammatic construction, ADC(3), and FRPA calculations~\cite{Barb07,Ortiz2013,Barb09}.

The remaining diagrams of Fig.~\ref{3ord} all include $3p2h$ and $3h2p$ configurations. 
These become necessary to reproduce the fragmentation patterns of shakeup configurations 
in particle removal and addition experiments, i.e. Dyson orbits beyond the main quasiparticle peaks. 
These contributions are computationally more demanding.
Diagrams~\ref{3ord_223B_1} to~\ref{3ord_332B_2} all describe interaction between $2p1h$~($2h1p$) and $3p2h$~($3h2p$) 
configurations. These are split into four contributions arising from two effective 2BFs and four that contain two irreducible 3B 
interactions.  Similarly, diagrams~\ref{3ord_323B_1} to~\ref{3ord_333B_3} are the first contributions 
to the configuration mixing among $3p2h$  or $3h2p$ states.

Appendix~\ref{app_Feyn} provides the Feynman diagram rules to compute the contribution associated with these
diagrams. Specific expressions for some diagrams in Fig.~\ref{3ord} are given.
We note that the Feynman rules remain unaltered whether one uses the original, $\h U$ and $\h V$,
or the effective, $\widetilde{U}$ or $\widetilde{V}$, interactions.  
Hence, symmetry factors due to equivalent lines remain unchanged.

\section{Equation of motion method}
\label{section3}

The perturbation theory expansion outlined in the previous section is useful to 
identify new contributions arising from the inclusion of 3B interactions. However, diagrams
up to third order alone do not necessarily incorporate all the necessary information to describe strongly
correlated quantum many-body systems. For example, the strong repulsive character
of the nuclear force at short distances requires explicit all-orders summations of 
ladder series. All-order summations of $2p1h$ and $2h1p$ are also required in finite
systems to achieve accuracy for the predicted ground state and separation energies, as well
as to preserve the correct analytic properties of the self-energy beyond second order.

To investigate approximation schemes for all-order summations including 3BFs, we now
develop the EOM method. This will provide special insight into 
possible self-consistent expansions of the irreducible self-energy, $\Sigma^\star$.
For 2B forces only, the EOM technique defines a hierarchy of equations that link each $n$-body GF 
to the $(n-1)$- and the $(n+1)$-body GFs. 
When extended to include 3BFs, the hierarchy also involves the $(n+2)$-body GFs. 
A truncation of this Martin-Schwinger hierarchy is necessary
to solve the system of equations \cite{Mar59} and can potentially give rise to
physically relevant resummation schemes.
Here, we will follow the footprints of Ref.~\cite{Mat71} and
apply truncations to obtain explicit equations for the 4-point (and 6-point, in the 3BF case) vertex functions. 

\subsection{Equation of motion for $G$ and proper self-energy}
\label{EOM_Sig}

The EOM for a given propagator is found by taking the derivative of its time arguments. The time
arguments are linked to the creation and annihilation operators in Eqs.~(\ref{G}) to 
(\ref{g6pt}) and hence the time dependence of these operators will drive that of the propagator \cite{Blaizot86}. 
The unperturbed 1B propagator can be written as the $n=0$ order term of Eq.~(\ref{gpert}),
\beq
\ii\hbar \, \gz_{\al \be}(t_\al - t_\be) = \lphizero\T[a^I_\al(t_\al){a_{\be}^I}^\dg(t_\be)]\rphizero \; , 
\label{G0}
\enq
Its time derivative will be given by the von Neumann 
equation for the operators in the interaction picture~\cite{Abr75}:
\beq
\ii\hbar\dev a^I_\al(t)= [a^I_\al(t),\h H_0] =  \vep^0_\al a^I_\al(t) \;. 
\label{eom_aI}
\enq
Taking the derivative of $\gz$ with respect to time and using Eq.~(\ref{eom_aI}), we find
\beq
\label{G0inv}
\left\{\ii\hbar\frac{\partial}{\partial t_\al}-\vep^0_\al\right\} G^{(0)}_{\al\be}(t_\al-t_\be)=\delta(t_\al-t_\be) \de_{\al\be} \, .
\enq
Note that the delta functions in time arise from the derivatives of the step-functions involved in the
time-ordered product. 

\begin{figure}[t!]
\begin{center}
\includegraphics[scale=0.5]{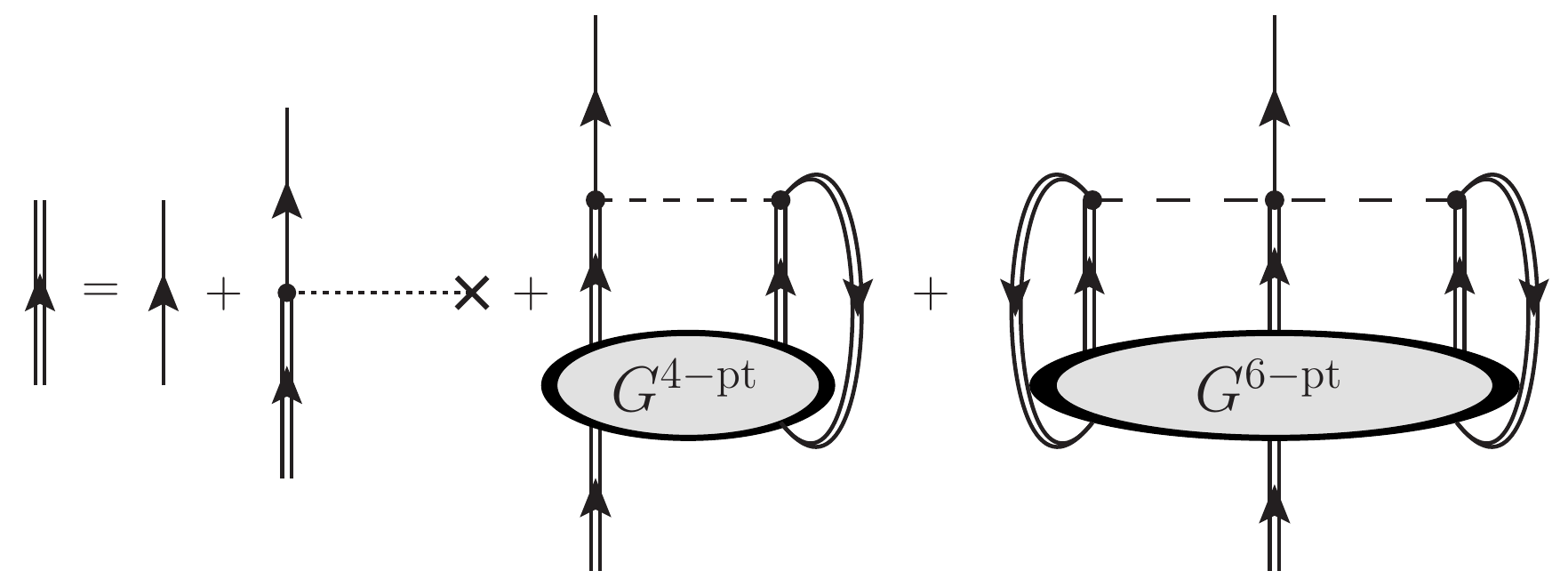}
\caption{Diagrammatic representation of the EOM, Eq.~(\ref{gexact}), 
for the dressed 1B propagator, $G$. The first term, given by a single line, defines the free 1B propagator, $\gz$.
The second term denotes the interaction with a bare 1B potential, 
whereas the third and the fourth terms describe interactions involving the intermediate propagation of two- 
and three-particle configurations. }
\label{eq_g}
\end{center}
\end{figure}

The same procedure applied to the exact 1B propagator, $G$, requires the
time-derivative of the operators in the Heisenberg picture. For the
original Hamiltonian of Eq.~(\ref{H}), the EOM for the annihilation operator reads:
\begin{widetext}
\beqn
\label{ader}
\ii\hbar\dev a_\al(t) = [a_\al(t),\h H]  &=&
\vep^0_\al a_\al(t) - \sum_{\de} U_{\al\de} a_\de(t) 
+ \f 1 2\sum_{\substack{\ep\\\de\mu}} V_{\al\ep,\de\mu} \ad\ep(t) a_\mu(t) a_\de(t) 
\\\nn &&
+\frac{1}{12}\sum_{\substack{\ep\ta\\\de\mu\lm}}
W_{\al\ep\ta,\de\mu\lm} \ad\ep(t)\ad\ta(t) a_\lm(t)a_\mu(t)a_\de(t)\,.
\enqn
This can now be used to take the derivative of the full 1B propagator in Eq.~(\ref{G}):
\beqn
\label{devG}
 \left\{\ii\hbar\frac{\partial}{\partial t_\al}-\vep^0_\al\right\}  G_{\al\be}(t_\al-t_\be) &=& \delta(t_\al-t_\be)\de_{\al\be} 
- \sum_{\de}  U_{\al\de} G_{\de \be}(t_\al-t_\be)
\\\nn 
&&
+\frac{1}{2} \sum_{\substack{\ep\\\de\mu}}  
V_{\al\ep,\de\mu} G^{4-{\rm pt}}_{\de\mu,\ep\be}(t_\al,t_\al;t_\al^+,t_\be) 
\\\nn &&  
+\frac{1}{12}\sum_{\substack{\ep\ta\\\de\mu\lm}}
W_{\al\ep\ta,\de\mu\lm} G^{6-{\rm pt}}_{\de\mu\lm,\ep\ta\be}(t_\al,t_\al,t_\al;t_\al^{++},t_\al^+,t_\be) \, .
\enqn
\end{widetext}
This equation links the 2-point GF to both the 4- and the 6-point GFs. Note that the connection with the latter is mediated 
by the 3BF and hence does not appear in the pure 2BF case. 
Regarding the time-arguments in Eq.~(\ref{devG}), the $t_\al^+$ and $t_\al^{++}$ present in the 4- and 6-point GFs
are necessary to keep the correct time-ordering in the creation operators when going from Eq.~(\ref{ader}) to Eq.~(\ref{devG}).
An analogous equation can be obtained for the derivative of the time argument $t_\be$. 
After some manipulation, involving the Fourier transforms of Eqs.~(\ref{Gmpt_ft}) and (\ref{G1B_ft}), 
one obtains the equation of motion for the SP 
propagator in frequency representation:
\begin{widetext}
\beqn
\label{gexact}
G_{\al\be}(\om) &=& G^{(0)}_{\al\be}(\om)
- \sum_{\ga\de}G^{(0)}_{\al\ga}(\om) U_{\ga\de} G_{\de\be}(\om)
\\\nn && 
-\frac{1}{2} \sum_{\substack{\ga\ep\\\de\mu}}G^{(0)}_{\al\ga}(\om) V_{\ga\ep,\de\mu}
\int \frac{d \om_1}{2 \pi} \int \frac{d \om_2}{2 \pi} 
G^{4-{\rm pt}}_{\de\mu,\be\ep}(\om_1,\om_2;\om,\om_1+\om_2-\om) 
\\\nn
&& +\frac{1}{12}\sum_{\substack{\ga\ep\ta\\\de\mu\lm}}G^{(0)}_{\al\ep}(\om)W_{\ga\ep\ta,\de\mu\lm}
\int \frac{d \om_1}{2 \pi}\int \frac{d \om_2}{2 \pi}\int \frac{d \om_3}{2 \pi}\int \frac{d \om_4}{2 \pi}
G^{6-{\rm pt}}_{\de\mu\lm,\ga\be\ta}(\om_1,\om_2,\om_3;\om_4,\om,\om_1+\om_2+\om_3-\om_4-\om)\,.
\enqn
\end{widetext}

Again, this involves both the 4- and the 6-point GFs, which appear due to the 2B and 3B interactions, respectively.
The equation now involves $n-2$ frequency integrals of the $n$-point GFs. 
The diagrammatic representation of this equation is given in Fig.~\ref{eq_g}.

The EOMs, Eqs.~(\ref{devG}) and~(\ref{gexact}), connect the 1B propagator to GFs of different orders. 
In general, starting from an $n$-body GF, the derivative of the time-ordering
operator generates a delta function between an incoming and outgoing particle, effectively separating a line and leaving
an ($n-1$)-body propagator. Conversely, the 2B part of the Hamiltonian introduces an extra pair of creation
and annihilation operators that adds another particle and leads to an ($n+1$)-body GF. 
For a 3B Hamiltonian, the ($n+2$)-body GF enters the EOM due to the commutator in Eq.~(\ref{ader}).
This implies that higher order GFs will be needed, at the same level of approximation, 
in the EOM hierarchy with 3BFs.

Eq.~(\ref{gexact}) gives an exact equation for the SP propagator $G$ that, however, requires the knowledge
of both the 4-point and 6-point GFs.  Full equations for the latter
require applying the EOMs to these propagators as well.  
Before that, however, it is possible to further simplify contributions in Eq.~(\ref{gexact}) by splitting the 
$n$-point GFs into two terms. 
The first one is relatively simple, involving the properly antisymmetrized 
independent propagation of $n$ dressed particles. The second term will
involve the interaction vertices, $\Gamma^{4-{\rm pt}}$ and $\Gamma^{6-{\rm pt}}$, 
1PI vertex functions that include all interaction effects \cite{Blaizot86}. 
These can be neatly connected to the irreducible self-energy.

\begin{figure}[t!]
\begin{center}
\includegraphics[scale=0.55]{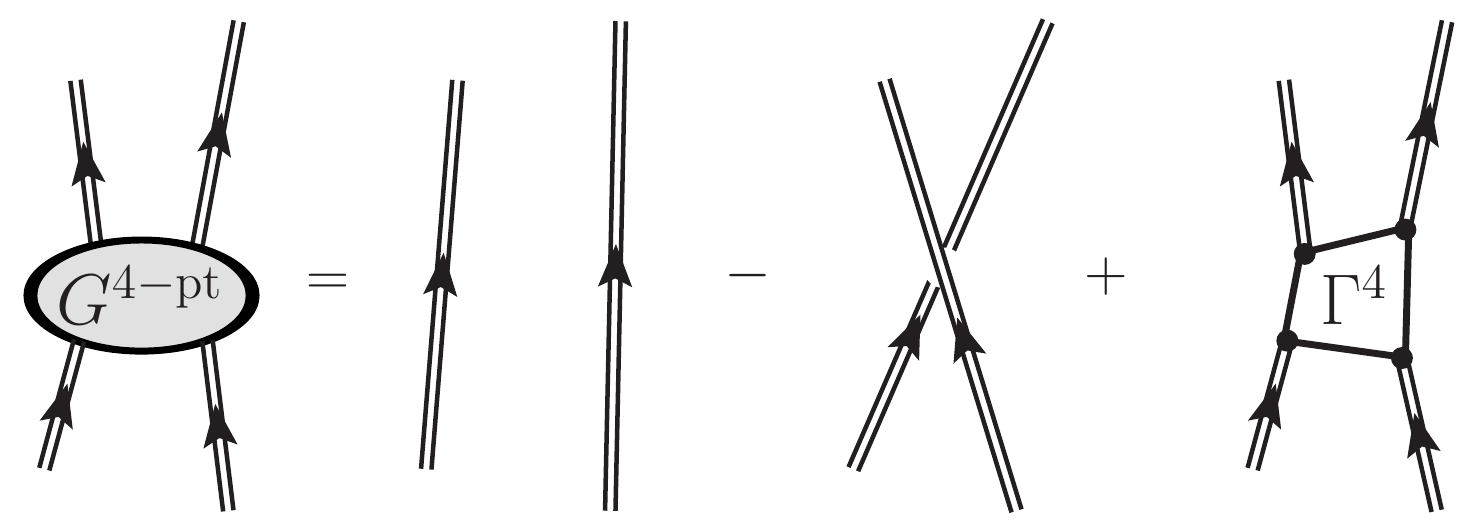}
\caption{Exact separation of the 4-point Green's function, $G^{4-{\rm pt}}$, in terms of non-interacting lines and a vertex function, as given in Eq.~(\ref{g4ptgamma}).
The first two terms are the direct and exchange propagation of two non-interacting and fully dressed particles. The last term defines the 4-point vertex function, $\Gamma^{4-{\rm pt}}$, involving the sum of all 1PI diagrams.}
\label{g4point}
\end{center}
\end{figure}

For the 4-point GF, this separation is shown diagrammatically in Fig.~\ref{g4point}. The first two terms involve two 
dressed fermion lines propagating independently, and their exchange as required by the Pauli principle. 
The remaining part, stripped of 
its external legs, can contain only 1PI diagrams which are collected in a vertex function, $\Gamma^{4-{\rm pt}}$. This is associated with interactions and, at lowest level, it would correspond to a 2BF. 
As we will see in the following, however, 3B interactions also provide contributions to $\Gamma^{4-{\rm pt}}$. 
The 4-point vertex function is defined by the following equation:
\begin{widetext}
\beqn
\label{g4ptgamma}
G^{4-{\rm pt}}_{\al\ga,\be\de}(\om_\al,\om_\ga;\om_\be,\om_\de) &=& 
 \ii \hbar \big[2\pi \de(\om_\al-\om_\be)G_{\al\be}(\om_\al)G_{\ga\de}(\om_\ga)
- 2\pi \de(\om_\ga-\om_\be)G_{\al\de}(\om_\al)G_{\ga\be}(\om_\ga)\big]
\\\nn &&  
+(\ii\hbar)^2\sum_{\substack{\ta\mu\\\nu\lm}}G_{\al\ta}(\om_\al)G_{\ga\mu}(\om_\ga)
\Gamma^{4-{\rm pt}}_{\ta\mu,\nu\lm}(\om_\al,\om_\ga;\om_\be,\om_\de)
G_{\nu\be}(\om_\be)G_{\lm\de}(\om_\de) \, .
\enqn 
\end{widetext}
Eq.~(\ref{g4ptgamma}) is \emph{exact} and is an implicit definition of $\Gamma^{4-{\rm pt}}$. 
Different many-body approximations arise when approximations are performed on this vertex function \cite{Dick04,Dick05}.

\begin{figure}[t!]
\begin{center}
\includegraphics[scale=0.6]{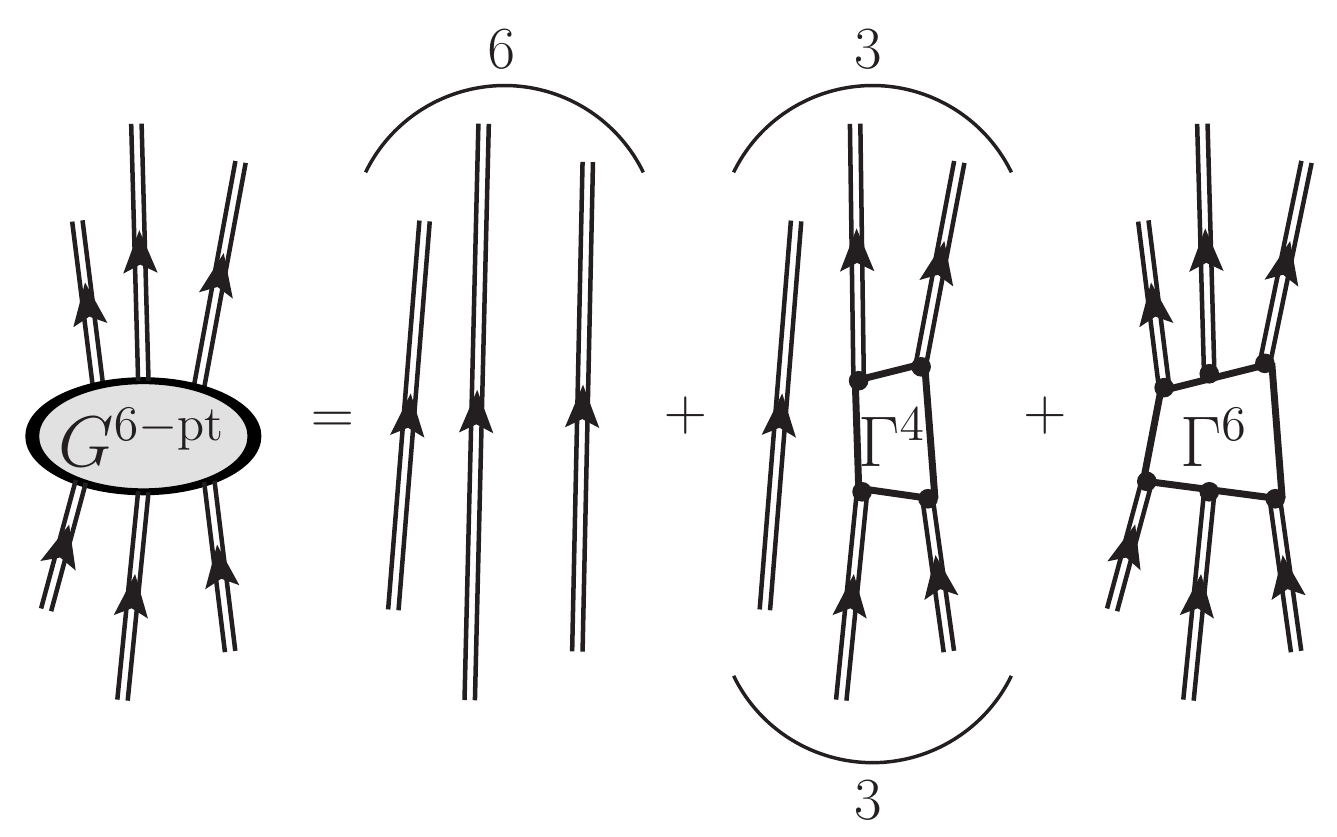}
\caption{Exact separation of the 6-point Green's function, $G^{6-{\rm pt}}$, in terms of 
non-interacting dressed fermion lines and vertex functions, as given in Eq.~(\ref{g6ptgamma}).
The first two terms gather non-interacting dressed lines and subgroups of 
interacting particles that are fully connected to each other. 
Round brackets with numbers above (below) these diagrams indicate the numbers of permutations of outgoing (incoming)
 legs needed to generate all possible diagrams.
The last term defines the 6-point 1PI vertex function~$\Gamma^{6-{\rm pt}}$.
}
\label{g6point}
\end{center}
\end{figure}

A similar expression holds for the 6-point GF. In this case, the diagrams that involve non interacting lines
can contain either all 3 dressed propagators moving independently from each other or groups of two lines interacting
through a 4-point vertex function. The remaining terms are collected
in a 6-point vertex function, $\Gamma^{6-{\rm pt}}$, which contains terms where all 3 lines are interacting. 
This separation is demonstrated diagrammatically in Fig.~\ref{g6point}.
The Pauli principle requires a complete antisymmetrization of these diagrams. 
For the ``free propagating" term, this implies all $3!=6$ permutations of the 3 lines. 
The second term, involving $\Gamma^{4-{\rm pt}}$, requires $3^2=9$ cyclic permutations 
within both incoming and outgoing legs. The 6-point vertex function is already antisymmetrized 
and hence no permutations are needed. 

The  equation corresponding to Fig.~\ref{g6point} is exact and provides an implicit definition of the 
$\Gamma^{6-{\rm pt}}$ vertex function:
\begin{widetext}
\beqn
\label{g6ptgamma}
G^{6-{\rm pt}}_{\al\ga\ep,\be\de\eta}&& (\om_\al,\om_\ga,\om_\ep;\om_\be,\om_\de,\om_\eta) =
\\\nn && 
(2\pi)^2 (\ii\hbar)^2\,{\cal A}_{[\{\al\om_\al\},\{\ga\om_\ga\},\{\ep\om_\ep\}]} \big[
\de(\om_\al-\om_\be)\,\de(\om_\ga-\om_\de)\, G_{\al\be}(\om_\al)G_{\ga\de}(\om_\ga)G_{\ep\eta}(\om_\ep)
 \big]
\\\nn && 
+ 2\pi(\ii\hbar)^3\, {\cal P}^{\rm cycl.}_{[\{\al\om_\al\},\{\ga\om_\ga\},\{\ep\om_\ep\}]}
                           {\cal P}^{\rm cycl.}_{[\{\be\om_\be\},\{\de\om_\de\},\{\eta\om_\eta\}]}
                           \Big[
                           \de(\om_\al-\om_\be) G_{\al\be}(\om_\al) \times
\\\nn && 
\qquad \qquad \qquad  \qquad \quad
\sum_{\substack{\ta\mu\\\nu\lm}}G_{\ga\ta}(\om_\ga)
G_{\ep\mu}(\om_\ep)\Gamma^{4-{\rm pt}}_{\ta\mu,\nu\lm}(\om_\ga,\om_\ep;\om_\de,\om_\eta) 
G_{\nu\de}(\om_\de)G_{\lm\eta}(\om_\eta) \Big]
\\\nn && 
+ (\ii\hbar)^4 
\sum_{\substack{\ta\mu\chi\\\nu\lm\xi}}G_{\al\ta}(\om_\al)G_{\ga\mu}(\om_\ga)
G_{\ep\chi}(\om_\ep)\Gamma^{6-{\rm pt}}_{\ta\mu\chi,\nu\lm\xi}(\om_\al,\om_\ga,\om_\ep;\om_\be,\om_\de, \om_\eta) G_{\nu\be}(\om_\be)G_{\lm\de}(\om_\de)G_{\xi\eta}(\om_\eta) \, .
\enqn 
\end{widetext}
Here, we have introduced the antisymmetrization operator, ${\cal A}$, which sums all possible permutations of pairs of indices and frequencies, $\{\al\om_\al\}$, with their corresponding sign.
Likewise, ${\cal P^{\rm cycl.}}$ sums all possible cyclic permutations of the index-frequency pairs. 
Again, let us stress that both $\Gamma^{4-{\rm pt}}$ and $\Gamma^{6-{\rm pt}}$ are formed of 1PI 
diagrams only, since they are defined by removing all external {\em dressed} legs from the $G^{4-{\rm pt}}$
and $G^{6-{\rm pt}}$ propagators.
However, they are still two-particle reducible, since they include diagrams that can be split by cutting two lines.
In general, $\Gamma^{4-{\rm pt}}$ and $\Gamma^{6-{\rm pt}}$ are solution of  all-orders summations analogous to the Bethe-Salpeter equation, in which the kernels are 2PI and 3PI vertices [see Eqs.~(\ref{GmLadd}) to
(\ref{GmParquet}) below].

\begin{figure}[t]
\begin{center}
\includegraphics[scale=0.5]{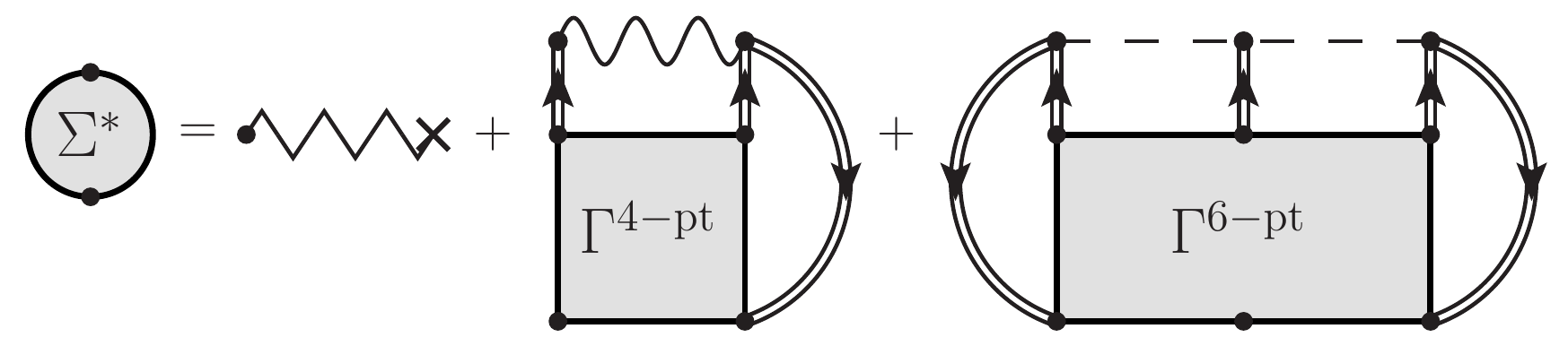}
\caption{Diagrammatic representation of the irreducible self-energy 
$\Sigma^\star$ by means of effective 1B and 2B potentials
and 1PI vertex functions, as given in Eq.~(\ref{irrself}).  
The first term is the energy independent part of $\Sigma^\star$ and contains all diagrams depicted in Fig.\ref{ueffective}.
The second and third terms are dynamical terms consisting of excited configurations generated through 2B and 3BFs.
This is an exact equation for Hamiltonians including 3BFs and it is not derived from perturbation theory. }
\label{self_en}
\end{center}
\end{figure}

Inserting Eqs.~(\ref{g4ptgamma}) and~(\ref{g6ptgamma}) into Eq.~(\ref{gexact}), 
and exploiting the effective 1B and 2B operators  defined in Eqs.~(\ref{ueff}) and (\ref{veff}),
one recovers the Dyson equation, Eq.~(\ref{Dyson}). One can therefore identify
the exact expression of the irreducible self-energy $\Sigma^\star$ in terms of 1PI 
vertex functions:
\begin{widetext}
\beqn
\label{irrself}
 \Sigma^\star_{\ga\de}(\om)&=& 
\widetilde U_{\ga\de} \\\nn && 
- \frac{(\ii\hbar)^2}{2}\sum_{\substack{\mu\\\nu\lm}}\sum_{\substack{\xi\ta\\\ep}}
\widetilde V_{\ga\mu,\nu\lm}\int\frac{\d\om_1}{2\pi}\int\frac{\d\om_2}{2\pi}
G_{\nu\xi}(\om_1)G_{\lm\ta}(\om_2)
\Gamma^{4-{\rm pt}}_{\xi\ta,\de\ep}(\om_1,\om_2;\om,\om_1+\om_2-\om)
G_{\ep\mu}(\om_1+\om_2-\om)
\\\nn &&
+\frac{(\ii\hbar)^4}{12}\sum_{\substack{\mu\phi\\\lm\nu\chi}}\sum_{\substack{\ta\xi\eta\\\ep\sig}}
W_{\mu\ga\phi,\lm\nu\chi}
\int\frac{\d\om_1}{2\pi}\int\frac{\d\om_2}{2\pi}\int\frac{\d\om_3}{2\pi}\int\frac{\d\om_4}{2\pi}
G_{\lm\ta}(\om_1)G_{\nu\xi}(\om_2)G_{\chi\eta}(\om_3) \times
\\\nn &&
\qquad \qquad
\Gamma^{6-{\rm pt}}_{\ta\xi\eta,\ep\de\sig}(\om_1,\om_2,\om_3;\om_4,\om,\om_1+\om_2+\om_3-\om_4-\om)
G_{\ep\mu}(\om_4)G_{\sig\phi}(\om_1+\om_2+\om_3-\om_4-\om)\,.
\enqn
\end{widetext}

The diagrammatic representation of Eq.~(\ref{irrself}) is shown in Fig.~\ref{self_en}.
We note that, as an irreducible self-energy, this should include all the
connected, 1PI diagrams. 
These can be regrouped in terms of \emph{skeleton} and \emph{interaction-irreducible} 
contributions, as long as $\Gamma^{4-{\rm pt}}$ and $\Gamma^{6-{\rm pt}}$ are expressed that way.
Note that effective interactions are used here. The interaction-reducible components of $\widetilde U$, $\widetilde{V}$ and $W$ 
are actually generated by contributions involving partially 
non-interacting propagators
contributions inside $G^{4-{\rm pt}}$ and $G^{6-{\rm pt}}$. 
The first two terms in both Eqs.~(\ref{g4ptgamma}) and~(\ref{g6ptgamma}) only
contribute to generate effective interactions. Note, however, that the 2B effective interaction
does receive contributions from both $\Gamma^{4-{\rm pt}}$ and $\Gamma^{6-{\rm pt}}$ in
the self-consistent procedure. 

The first term entering  Eq.~(\ref{irrself}) is the energy-independent contribution to
the irreducible self-energy, already found in Eq.~(\ref{eq:1ord}). 
This includes the subtraction of the auxiliary field, $\hat{U}$, as well as the 1B interaction-irreducible
contributions due to the 2B and 3BFs. 
Once again, we note that the definition of this term, shown in Fig.~\ref{ueffective}, involves fully correlated
density matrices. Consequently, even though this is a static contribution, it goes beyond the 
Hartree-Fock approximation.
The dispersive part of the self-energy is described by the second and third terms on the right-side
of Eq.~(\ref{irrself}). These account for all higher-order contributions
and incorporate correlations on a 2B and 3B level associated with the vertex functions
$\Gamma^{4-{\rm pt}}$ and $\Gamma^{6-{\rm pt}}$, respectively. 
In Sec.~\ref{Sig_vs_gamma} below, we will expand these vertices up to second order and 
show that Eq.~(\ref{irrself}) actually generates all diagrams  derived in Sec.~\ref{sec_PT_3ord}. 


\subsection{Equation of motion for $G^{4-{\rm pt}}$ and $\Gamma^{4-{\rm pt}}$}
\label{EOM_GII}

\begin{figure}[t]
\begin{center}
\includegraphics[scale=0.55]{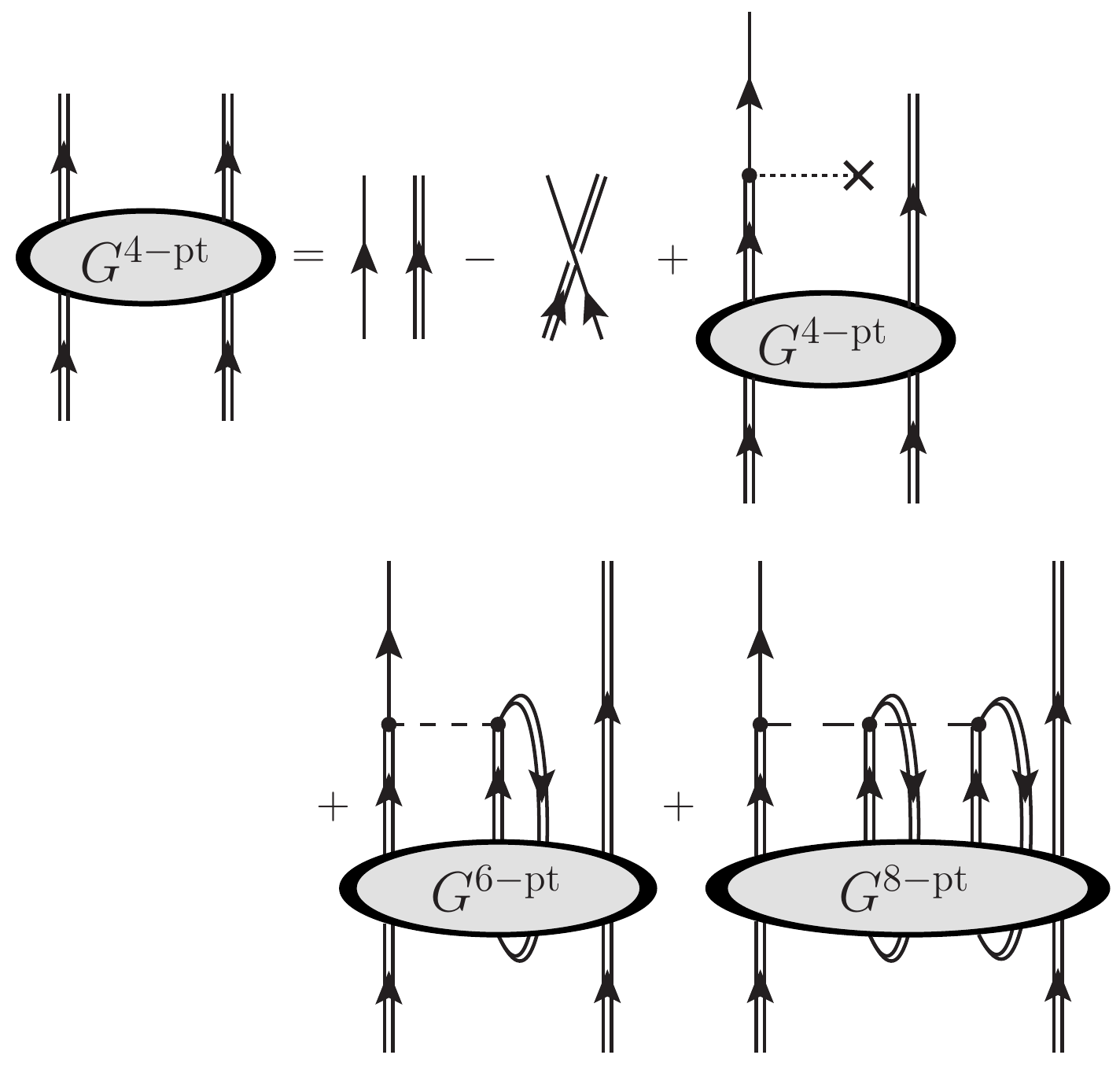}
\caption{Diagrammatic representation of the EOM for the four-point propagator, $G^{4-{\rm pt}}$, given in Eq.~(\ref{gIIexact}).
The last term, involving an 8-point GF, arises due to the presence of 3B interactions. }
\label{eq_g2}
\end{center}
\end{figure}

We now apply the EOM method to the 4-point GF. This will provide insight into approximation schemes that 
involve correlations at or beyond the 2B-level. Let us stress that our final aim is to obtain generic
nonperturbative approximation schemes in the many-body sector.
Taking the time derivative of the first argument in Eq.~(\ref{g4pt}) and following the same 
procedure as in Sec.~\ref{EOM_Sig}, we find:
\begin{widetext}
\beqn
\label{gIIexact}
G^{4-{\rm pt}}_{\al\ga,\be\de}(\om_\al,\om_\ga;\om_\be,\om_\de) &=& 
\,\ii\hbar\,[ 2 \pi \de(\om_\al-\om_\be) G^{(0)}_{\al\be}(\om_\al) G_{\ga\de}(\om_\ga)-
2 \pi \de(\om_\ga-\om_\be) G^{(0)}_{\al\de}(\om_\al)G_{\ga\be}(\om_\ga)]
\\\nn && 
+ \sum_{\mu\lm}G^{(0)}_{\al\mu}(\om_\al) U_{\mu\lm} G^{4-{\rm pt}}_{\lm\ga,\be\de}(\om_\al,\om_\ga;\om_\be,\om_\de)
\\\nn &&
-\frac{1}{2} \sum_{\substack{\mu\ep\\\lm\ta}}G^{(0)}_{\al\mu}(\om_\al)
 V_{\mu\ep,\lm\ta}
\int\frac{d \om_1}{2 \pi}\int\frac{d \om_2}{2 \pi}
G^{6-{\rm pt}}_{\lm\ta\ga,\be\ep\de}(\om_1,\om_2,\om_\ga;\om_\be,\om_1+\om_2-\om_\al,\om_\de) 
\\\nn&&
+\frac{1}{12}\sum_{\substack{\mu\ep\chi\\\lm\ta\eta}}G^{(0)}_{\al\mu}(\om_\al)
W_{\mu\ep\chi,\lm\ta\eta}
\int\frac{d \om_1}{2 \pi}\int\frac{d \om_2}{2 \pi}\int\frac{d \om_3}{2 \pi}\int\frac{d \om_4}{2 \pi}
\\ && \quad\quad\quad\quad
G^{8-pt}_{\lm\ta\eta\ga,\be\ep\chi\de}(\om_1,\om_2,\om_3,\om_\ga;\om_\be,\om_4,\om_1+\om_2+\om_3-\om_\al-\om_4,\om_\de)\,. \nn
\enqn
\end{widetext}
which is the analogous of Eq.~(\ref{gexact}) for the SP propagator.
As expected, the EOM connects the 2-body (4-point) GFs to other propagators. 
The 1B propagator term just provides the non-interacting dynamics, with the proper antisymmetrization.
The interactions bring in admixtures with the 4-point GFs itself, via the one-body potential, but also
with the 6- and 8-point GFs, via the the 2B and the 3B interactions, respectively. Similarly to what 
we observed in Eq.~(\ref{gexact}), the dynamics involve $n-4$ frequency integrals of the $n$-point GFs. 
The diagrammatic representation of this equation is given in Fig. \ref{eq_g2}.

\begin{figure}[t!]
\begin{center}
\includegraphics[scale=0.6]{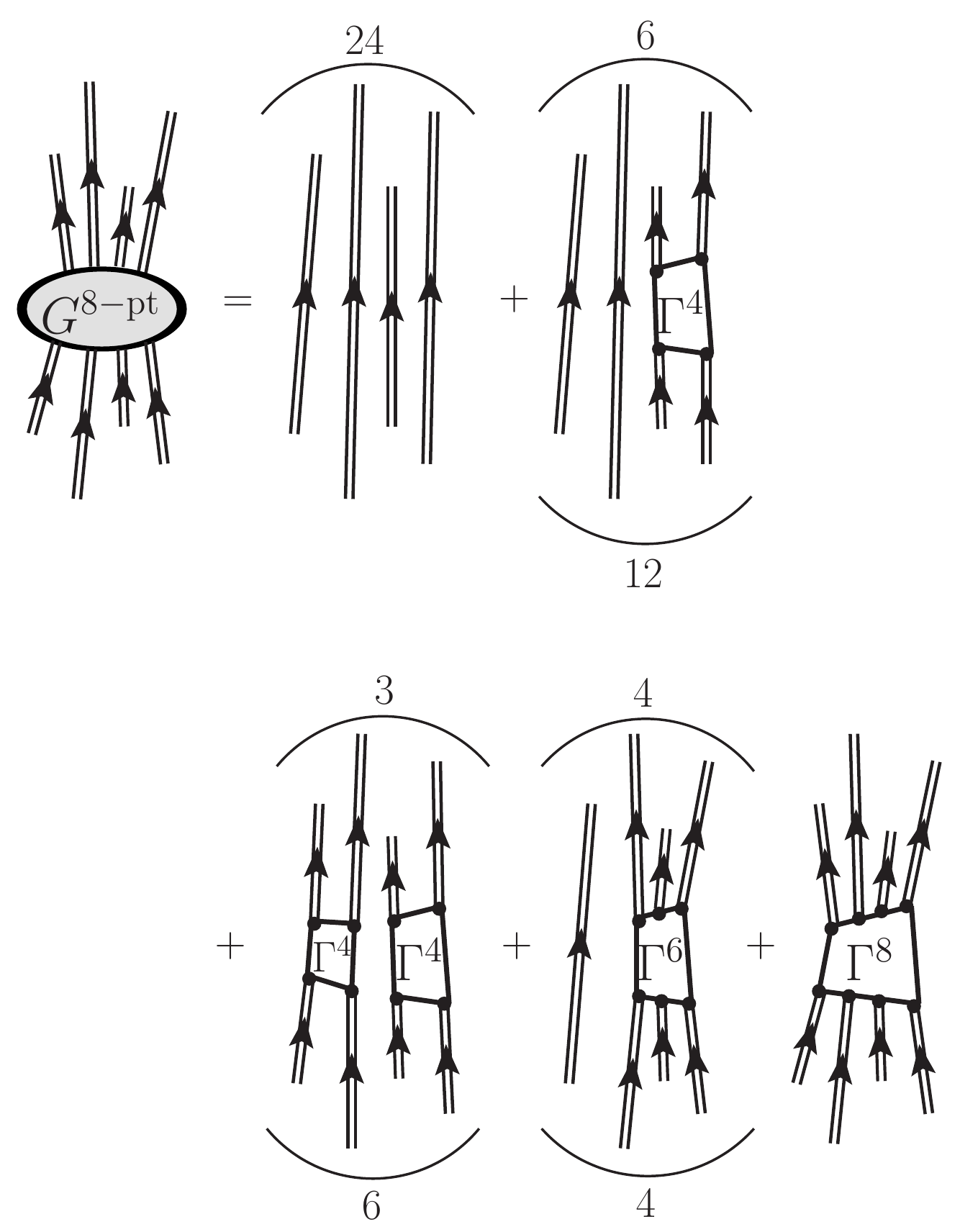}
\caption{Exact separation of the 8-point Green's function, $G^{8-{\rm pt}}$, in terms of 
non-interacting lines and vertex functions.
The first four terms gather non-interacting dressed lines and subgroups of 
interacting particles that are fully connected to each other. 
Round brackets with numbers above (below) these diagrams indicate the numbers of permutations of outgoing (incoming)
legs needed to generate all possible diagrams.
The last term defines the 8-point 1PI vertex function~$\Gamma^{8-{\rm pt}}$.
\label{g8point} }
\end{center}
\end{figure}

To proceed further, we follow the steps of the previous section and of Ref.~\cite{Mat71} 
and split the 8-point GF into free dressed propagators and 1PI vertex functions.
This decomposition is shown in Fig.~\ref{g8point}. 
In addition to the already-defined vertex functions, one needs 1PI objects with 4 incoming and
outgoing indices. To this end, we introduce the 8-point
vertex $\Gamma^{8-{\rm pt}}$ in the last term. 
Note that due care has to be taken
of all antisymmetrization possibilities when groups of fermion lines that are not
connected by $\Gamma^{8-{\rm pt}}$ are considered. 
The first term, for instance, involves 4 non-interacting but dressed fermion lines, and there
are $4!=24$ possible combinations. 
There are $\binom{4}{2}\binom{4}{2}\f 1 2=72$ equivalent terms involving two non-interacting lines and 
a single $\Gamma^{4-{\rm pt}}$, as in the second term 
of Fig.~\ref{g8point}. The double $\Gamma^{4-{\rm pt}}$ contribution (third term) can be 
obtained in $6 \times 3=18$ equivalent ways.

\begin{figure*}
\centering
\includegraphics[scale=0.46]{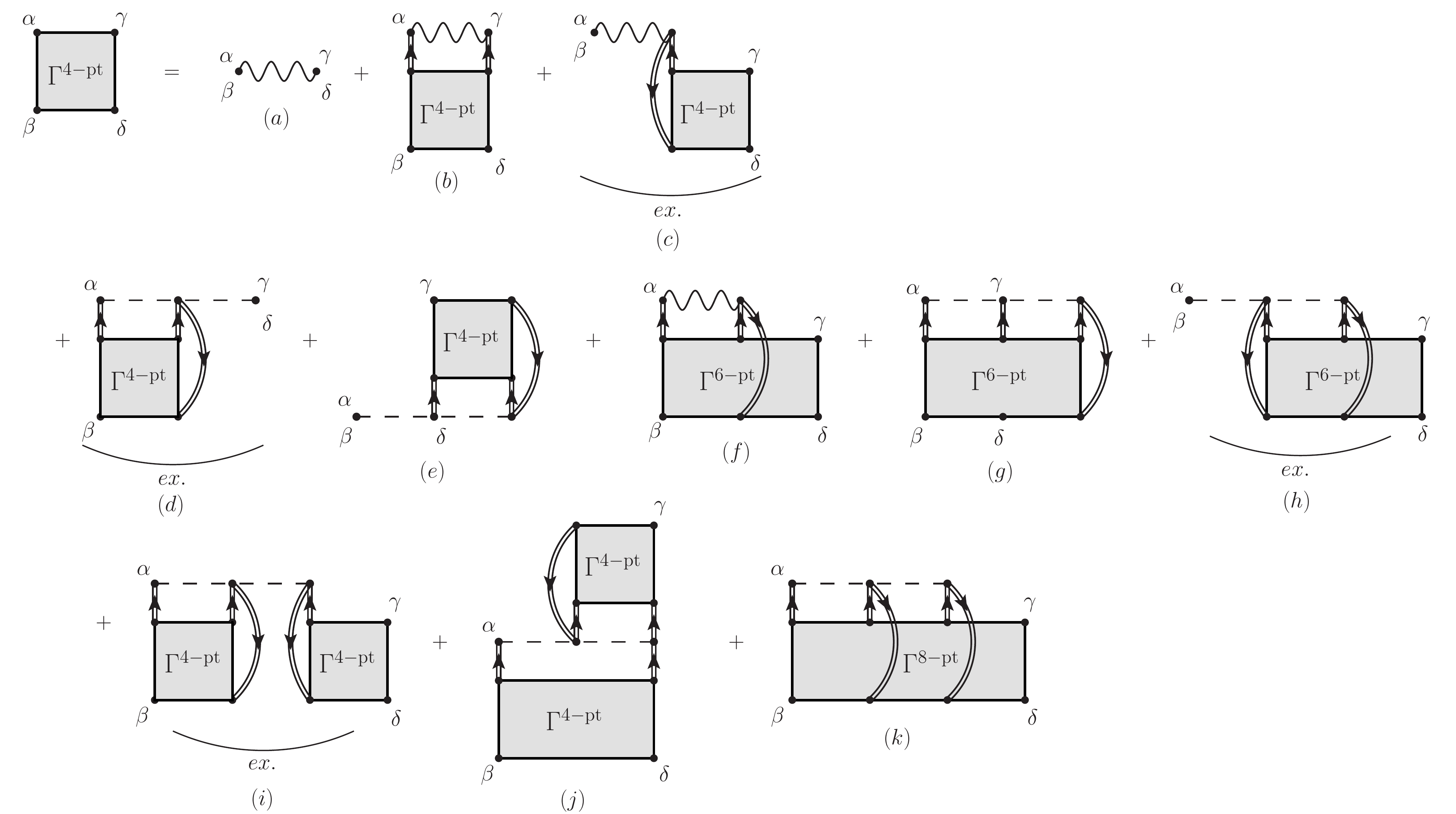}
\caption{Self-consistent expression for the $\Gamma^{4-{\rm pt}}$ vertex function derived from 
the EOM for $G^{4-{\rm pt}}$.
The round brackets underneath some of the diagrams indicate that the term obtained by 
exchanging the $\{\be\om_\be\}$ and $\{\de\om_\de\}$ arguments must also be included. 
Diagrams (a), (b), (c) and (f) are the only ones present for 2B Hamiltonians, although (f) also contains some intrinsic 3BF contributions
such as the $\{\al\om_\al\} \leftrightarrow \{\ga\om_\ga\}$ exchange of (e). All other diagrams 
arise from the inclusion of 3B interactions.  
Diagram (b) is responsible for generating the ladder summation,  the direct part of (c) generates the series of 
antisymmetrized rings, and the three sets together [(b), (c) and the exchange of (c)] would give rise to a Parquet-type 
resummation.}
\label{gamma4pt}
\end{figure*}

With this decomposition at hand, one can now proceed and find an equation for the 
4-point vertex function, $\Gamma^{4-{\rm pt}}$. 
Inserting the exact decompositions of the 4-, 6- and 8-point GFs, given respectively by 
Figs.~\ref{g4point}, \ref{g6point} and~\ref{g8point}, 
into the EOM, Eq.~(\ref{gIIexact}), one obtains an equation with $\Gamma^{4-{\rm pt}}$
on both sides. The diagrammatic representation of this self-consistent equation is
shown in Fig.~\ref{gamma4pt}.

A few comments are in order at this point.
The left hand side of Eq.~(\ref{gIIexact}) in principle contains two dressed and non interacting 
propagators, as shown in the first two terms of Fig.~\ref{g4point}. 
In the right hand side, however, one of the 1B propagators is not dressed. 
When expanding the GFs in
Eq.~(\ref{gIIexact}) in terms of the $\Gamma^{2n-{\rm pt}}$ vertex functions,  the remaining 
contributions to the Dyson equation arise automatically (Fig.~\ref{eq_g}). 
The free unperturbed line, therefore, becomes dressed. As a consequence, the
pair of dressed non-interacting propagators cancel out exactly on both sides of 
Eq.~(\ref{gIIexact}).  
This dressing procedure of the $\gz$ propagator happens only partially in the last three terms 
of the equation and has been disregarded in our derivation. In this sense,
Fig.~\ref{gamma4pt} should be taken as an approximation to the exact EOM
 for $G^{4-{\rm pt}}$.

Eq.~(\ref{gIIexact})  links  1B, 2B, 3B and 4B propagators. 
Correspondingly, Fig.~\ref{gamma4pt} involves higher order vertex functions, such as 
$\Gamma^{6-{\rm pt}}$ and $\Gamma^{8-{\rm pt}}$, which are in principle coupled, 
through their own EOMs, to more complex GFs. 
The hierarchy of these equations has to be necessarily truncated. 
In Ref.~\cite{Mat71}, truncation schemes were explored 
by neglecting the $\Gamma^{6-{\rm pt}}$ vertex function 
at the level of Fig.~\ref{gamma4pt} ($\Gamma^{8-{\rm pt}}$ did not appear in the 2BF-only case). 
This level of truncation is already sufficient to retain physically-relevant 
subsets of diagrams, such as ladders and rings. Let us note, in particular, that the 
summation of these infinite series leads to nonperturbative many-body schemes. 
For completeness, we show in Fig.~\ref{gamma4pt} all contributions coming also from the $\Gamma^{6-{\rm pt}}$ and $\Gamma^{8-{\rm pt}}$ vertices, 
many of them arising from 3BFs. 

We have ordered the diagrams in Fig.~\ref{gamma4pt} in terms of
increasing contributions from 3BFs and in the order of perturbation
theory at which they start contributing to $\Gamma^{4-{\rm pt}}$. 
Intuitively, we expect that this should order them in decreasing importance. 
Diagrams \ref{gamma4pt}a, \ref{gamma4pt}b, \ref{gamma4pt}c and \ref{gamma4pt}f  
are those that are also present in the 2BF-only case. 
Diagram \ref{gamma4pt}f, however, is of a mixed nature: it can contribute only at third order with 
effective 2BFs, but does contain interaction-irreducible 3BF contributions at second order 
that are similar to diagrams~ \ref{gamma4pt}d and  \ref{gamma4pt}e.
Diagrams \ref{gamma4pt}d-h all contribute to $\Gamma^{4-{\rm pt}}$ at second order,
although the first three require a combination of a $\widetilde{V}$ and a $W$ term.
The remaining diagrams in this group, \ref{gamma4pt}g and \ref{gamma4pt}h, require
two 3B interactions at second order and are expected to be subleading.
Note that \ref{gamma4pt}d is antisymmetric in $\al$ and $\ga$, but it must also 
be antisymmetrized with respect to $\be$ and $\de$. 
Its conjugate contribution, \ref{gamma4pt}e, should not be further antisymmetrized
in $\al$ and $\ga$,
because such exchange term is already included in \ref{gamma4pt}f.
All the remaining terms, \ref{gamma4pt}i-k, only contribute from the third order on.

The simplest truncation schemes to $\Gamma^{4-{\rm pt}}$ come from considering 
the first three terms of Fig.~\ref{gamma4pt}, which involve effective 2BFs only. In the pure 2B case, these have already been discussed in the literature \cite{Mat71}.
Retaining diagrams \ref{gamma4pt}a and \ref{gamma4pt}b leads to the ladder resummation used in recent studies of infinite nucleonic matter~\cite{Soma08, Car13}:
\begin{widetext}
\beqn
\label{GmLadd}
&&\Gamma^{4_\text{ladd}}_{\al\ga,\be\de}(\om_\al,\om_\ga;\om_\be,\om_\al+\om_\ga-\om_\be)=
\widetilde V_{\al\ga,\be\de} 
\\\nn && \quad\quad\quad
+ \frac{\ii\hbar}{2}\int\frac{\d\om_1}{2\pi}\sum_{\ep\mu\ta\lm}
\widetilde V_{\al\ga,\ep\mu}G_{\ep\ta}(\om_1)G_{\mu\lm}(\om_\al+\om_\ga-\om_1)
\Gamma^{4_\text{ladd}}_{\ta\lm,\be\de}(\om_1,\om_\al+\om_\ga-\om_1;\om_\be,\om_\al+\om_\ga-\om_\be) \; ,
\enqn
where we have explicitly used the fact that $\Gamma^{2n-{\rm pt}}$ is only defined 
when incoming and outgoing energies are conserved. 
Likewise, diagrams \ref{gamma4pt}a and the direct contribution of \ref{gamma4pt}c generate a series of ring diagram which correspond to the antisymmetrized version of the random phase approximation (RPA): 
\beqn
\label{GmRing}
&&\Gamma^{4_\text{ring}}_{\al\ga,\be\de}(\om_\al,\om_\ga;\om_\be,\om_\al+\om_\ga-\om_\be)=
\widetilde V_{\al\ga,\be\de} 
\\\nn &&\quad\quad\quad
- \ii\hbar\int\frac{\d\om_1}{2\pi}\sum_{\ep\mu\ta\lm} 
\widetilde V_{\al\ep,\be\mu}G_{\mu\lm}(\om_1)G_{\ta\ep}(\om_1-\om_\al+\om_\be)
\Gamma^{4_\text{ring}}_{\lm\ga,\ta\de}(\om_1,\om_\ga;\om_1-\om_\al+\om_\be,\om_\al+\om_\ga-\om_\be)
 \, .
\enqn
Adding up the first three contributions together, \ref{gamma4pt}a-c, and 
including the exchange, will generate a Parquet-type of resummation,
with ladders and rings embedded into each other:
\beqn
\label{GmParquet}
&&\Gamma^{4_\text{Parquet}}_{\al\ga,\be\de}(\om_\al,\om_\ga;\om_\be,\om_\al+\om_\ga-\om_\be)=
\widetilde V_{\al\ga,\be\de} 
\\\nn && \quad\quad\quad
+ \ii\hbar\int\frac{\d\om_1}{2\pi}\sum_{\ep\mu\ta\lm}
\left[ \frac{1}{2}\widetilde V_{\al\ga,\ep\mu}G_{\ep\ta}(\om_1)G_{\mu\lm}(\om_\al+\om_\ga-\om_1)
\Gamma^{4_\text{Parquet}}_{\ta\lm,\be\de}(\om_1,\om_\al+\om_\ga-\om_1;\om_\be,\om_\al+\om_\ga-\om_\be) \right.
\\ \nn && 
\qquad\qquad\qquad\qquad\qquad \left.
- \widetilde V_{\al\ep,\be\mu}G_{\mu\lm}(\om_1)G_{\ta\ep}(\om_1-\om_\al+\om_\be)
\Gamma^{4_\text{Parquet}}_{\lm\ga,\ta\de}(\om_1,\om_\ga;\om_1-\om_\al+\om_\be,\om_\al+\om_\ga-\om_\be)\right.
\\\nn &&
\qquad\qquad\qquad\qquad\qquad \left.
+\widetilde V_{\al\ep,\de\mu}G_{\mu\lm}(\om_1)G_{\ta\ep}(\om_1+\om_\ga-\om_\be)
\Gamma^{4_\text{Parquet}}_{\lm\ga,\ta\be}(\om_1,\om_\ga;\om_1+\om_\ga-\om_\be,\om_\be)\right]   \; .
\enqn
\end{widetext}
Eqs.~(\ref{GmLadd}) and~(\ref{GmRing}) can be solved in a more or less simple fashion
because the corresponding vertex functions effectively depend on only one frequency
($\Omega=\om_\al+\om_\ga$ and $\Omega=\om_\al-\om_\be$, respectively). Hence 
these two resummation schemes have been traditionally used to study 
extended systems~\cite{Dick04,Arya1998,Onida2002}. 
The simultaneous resummation of both rings and ladders within the self-energy is 
possible for finite systems, and it is routinely used in both quantum chemistry and 
nuclear physics~\cite{Schirmer1983,Danovich2011,Barb07,Cipol13}.
The Parquet summation, as shown in Eq.~(\ref{GmParquet}), does require all three 
independent frequencies and it is difficult to implement numerically. 
Specific approximations to rewrite these in terms of two-time vertex functions have been 
recently attempted~\cite{Bergli2011}, but further developments are still required. 

The next approximation to $\Gamma^{4-{\rm pt}}$ would involve diagrams \ref{gamma4pt}d, \ref{gamma4pt}e, and
the exchange part included in \ref{gamma4pt}f. All these should be added together to preserve the antisymmetry
and conjugate properties of the vertex function. The resulting contributions still depend on all three frequencies and 
cannot be simply embedded in all-order summations such as the ladder, Eq.~(\ref{GmLadd}), or the ring, 
Eq.~(\ref{GmRing}), approximations. However, these diagrams could be used to obtain corrections, at first order in the interaction-irreducible $\hat W$, to the previously calculated 4-point vertices. The explicit expression for these 
terms is:
\begin{widetext}
\beqn
\label{DeltaGm}
&&\Delta\Gamma^{4_{d+e+e'}}_{\al\ga,\be\de}(\om_\al,\om_\ga;\om_\be,\om_\al+\om_\ga-\om_\be)=
\frac{(\ii\hbar)^2}{2}\int\frac{\d\om_1}{2\pi}\int\frac{\d\om_2}{2\pi}
\sum_{\substack{\ep\mu\xi\\ \ta\lm\nu}}
\\\nn && \quad\quad
\left[- W_{\al\nu\ga,\ep\mu\de} \, G_{\ep\ta}(\om_1)G_{\mu\lm}(\om_2) \, 
\Gamma_{\ta\lm,\be\xi}(\om_1,\om_2; \om_\be,\om_1+\om_2-\om_\be) \, G_{\xi\nu}(\om_1+\om_2-\om_\be)\right.
\\\nn && \left.  \quad\quad
+ W_{\al\nu\ga,\ep\mu\be} \, G_{\ep\ta}(\om_1)G_{\mu\lm}(\om_2) \, 
\Gamma_{\ta\lm,\de\xi}(\om_1,\om_2;\om_\al+\om_\ga-\om_\be, \om_1+\om_2-\om_\al-\om_\ga+\om_\be) \,
G_{\xi\nu}(\om_1+\om_2-\om_\al-\om_\ga+\om_\be)\right.
\\\nn && \left. \quad\quad
- \Gamma_{\ga\nu,\ep\mu}(\om_\ga,\om_1+\om_2-\om_\ga;\om_1,\om_2) \, 
G_{\ep\ta}(\om_1)G_{\mu\lm}(\om_2)   \,  W_{\al\ta\lm,\be\de\xi} \,
G_{\xi\nu}(\om_1+\om_2-\om_\ga)\right.
\\\nn && \left. \quad\quad
+ \Gamma_{\al\nu,\ep\mu}(\om_\al,\om_1+\om_2-\om_\al;\om_1,\om_2) \, 
G_{\ep\ta}(\om_1)G_{\mu\lm}(\om_2)   \,  W_{\ga\ta\lm,\be\de\xi} \,
G_{\xi\nu}(\om_1+\om_2-\om_\al)\right] \; .
\enqn
\end{widetext}
Eq.~(\ref{DeltaGm}) has some very attractive features. First, it should provide the 
dominant contribution beyond those associated with the effective 2B interaction, 
$\widetilde{V}$. Perhaps more importantly, this contribution 
can be easily calculated in terms of one of the two-time vertex functions,
 $\Gamma^{4_{ladd}}$ and~$\Gamma^{4_{ring}}$. 
This could then be inserted in Eq.~(\ref{g4ptgamma}) to generate corrections of
expectation values of 2B operators stemming from purely irreducible 3B contributions.
A similar correction for the irreducible self-energy is also discussed in the next section.

Once a truncation scheme is chosen at the level of the vertex functions, one can
immediately derive a diagrammatic approximation for the self-energy \cite{Dick05}. 
Conserving approximations can plausibly be derived from some of these truncation schemes
\cite{Baym:1961zz}. A general derivation of $\Phi$-derivability with 3BF should be possible,
but goes beyond the scope of this work.

\subsection{Contributions to the irreducible self-energy}
\label{Sig_vs_gamma}

\begin{figure}[t!]
\begin{center}
\includegraphics[scale=0.46]{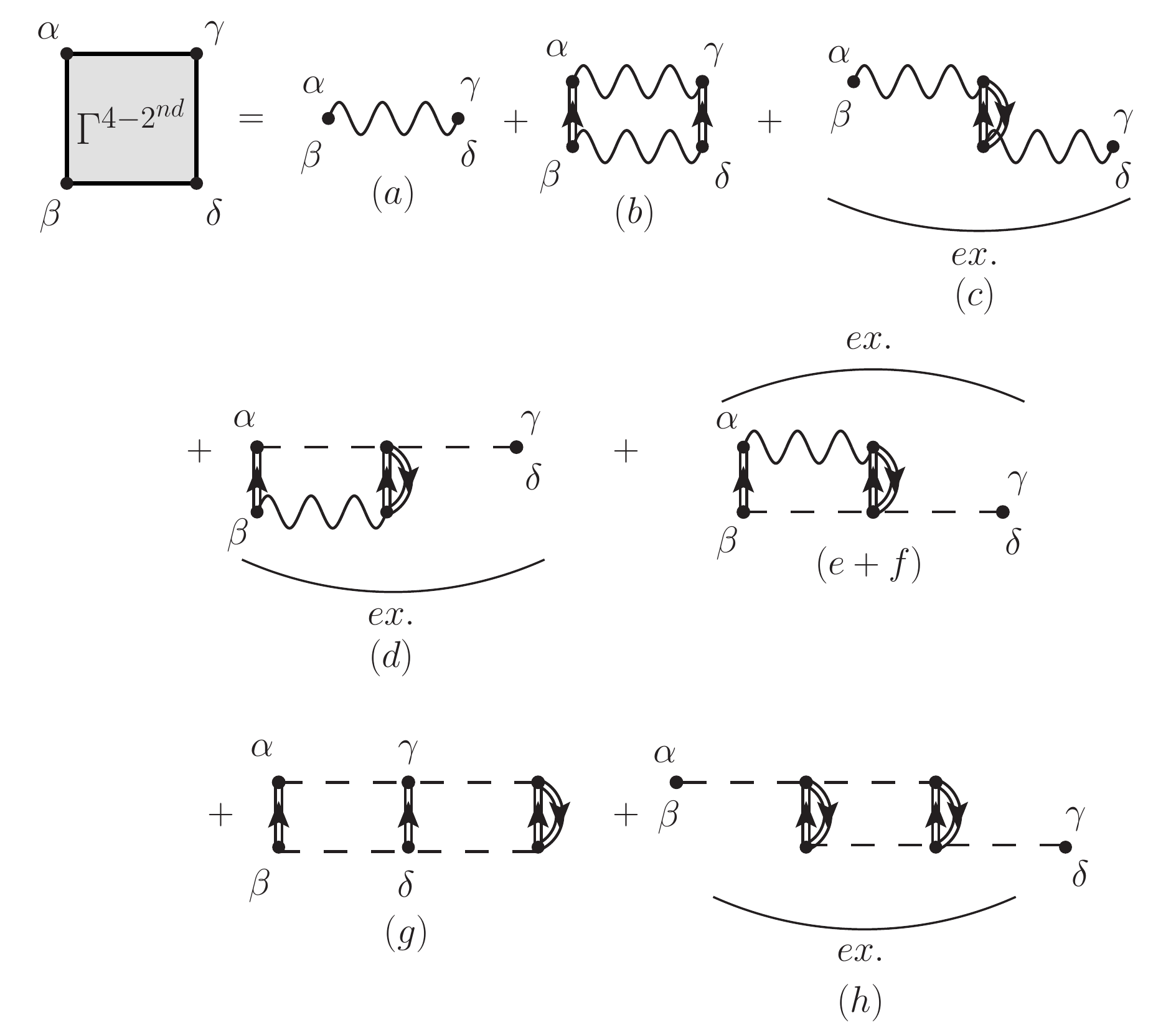}
\caption{Skeleton and interaction-irreducible diagrams contributing to the
 $\Gamma^{4-{\rm pt}}$ vertex function up to second order.
The round brackets above (below) some diagrams indicate that the exchange diagram
between the $\{\al\om_\al\}$ and $\{\ga\om_\ga\}$ \hbox{($\{\be\om_\be\}$ and $\{\de\om_\de\}$)}
arguments must also be included. }
\label{g4pt_2nd}
\end{center}
\end{figure}

In this subsection, we demonstrate the correspondence between the techniques derived in 
Section~\ref{section2} and the EOM method. In particular, we want to show how
 the perturbative expansion of Eq.~(\ref{gpert}) leads to the self-energy
obtained with the EOM expression, Eq.~(\ref{irrself}). We will do this by expanding the 
self-energy up to third order and showing the equivalence of both approaches at this order.  
To this end, we need to expand  the vertex functions in terms of the 
effective Hamiltonian, $\widetilde{H}_1$.
The lowest order terms entering $\Gamma^{4-{\rm pt}}$ can be easily read 
from Fig.~\ref{gamma4pt}.  
We show these second-order, skeleton and interaction-irreducible diagrams in 
Fig.~\ref{g4pt_2nd}. Only the first three terms would contribute for a 2BF. There are
two terms involving mixed 2BFs and 3BFs, whereas the final two contributions come from
two independent 3BFs. Note that, to get the third order expressions of the self-energy,
we expand the vertex functions to second order, i.e. one order less. 

\begin{figure}[t!]
\begin{center}
\includegraphics[scale=0.45]{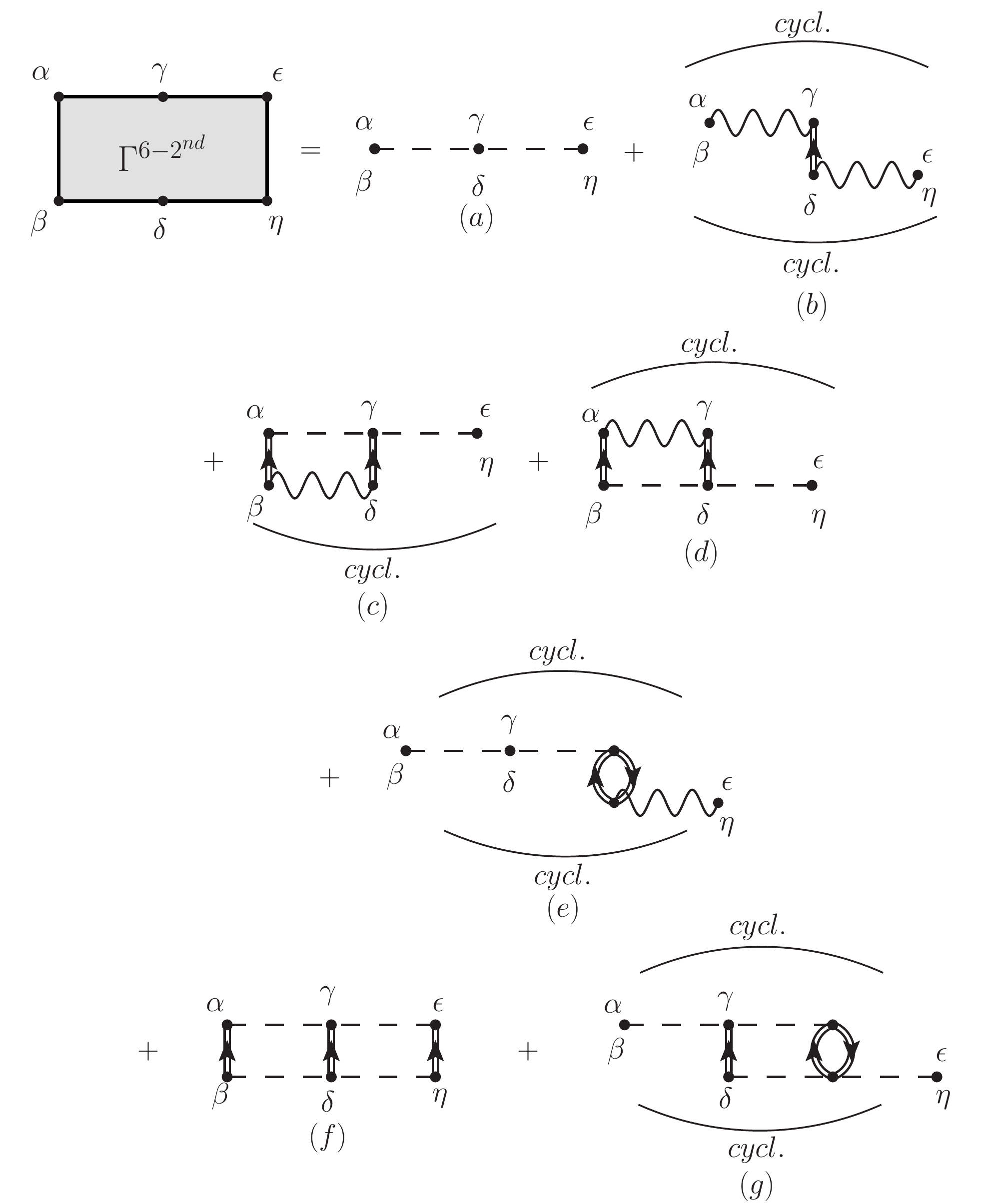}
\caption{ The same as Fig.~\ref{g4pt_2nd} for the  $\Gamma^{6-{\rm pt}}$ vertex function.
The round brackets above (below) some diagrams indicate that cyclic permutations of the $\{\al\om_\al\}$, $\{\ga\om_\ga\}$ and $\{\ep\om_\ep\}$ ($\{\be\om_\be\}$, $\{\de\om_\de\}$ and $\{\eta\om_\eta\}$) arguments must also be included. }
\label{g6pt_2nd}
\end{center}
\end{figure}

Analogously, we display the expansion up to second order of $\Gamma^{6-{\rm pt}}$ in 
Fig.~\ref{g6pt_2nd}. 
Most contributions to this vertex function contain 3BFs.
The lowest order term, for instance, is given by the 3B interaction itself. 
Note, however, that second 
order terms formed of 2B effective interactions are possible, such as the second term on the 
right hand side of Fig.~\ref{g6pt_2nd}. These will eventually be connected with a 3BF to give a mixed self-energy
contribution [see Eq.~(\ref{irrself}) and Fig.~\ref{self_en}]. 

If one includes the diagrams in Figs.~\ref{g4pt_2nd} and \ref{g6pt_2nd} into
the irreducible self-energy $\Sigma^\star$ of Fig.~\ref{self_en}, 
all the diagrams discussed in Eq.~\ref{eq:1ord}, Fig.~\ref{2ord} and Fig.~\ref{3ord} 
of Sec.~\ref{sec_PT_3ord} are recovered. 
This does prove, at least up to third order, the correspondence between the 
perturbative expansion approach and the EOM method for the GFs. 
Proceeding in this manner to higher orders, one should obtain equivalent diagrams 
all the way through. 

It is important to note that diagrams representing conjugate contributions to $\Sigma^\star$ 
are generated by different, not necessarily conjugate, terms 
of $\Gamma^{4-{\rm pt}}$ and~$\Gamma^{6-{\rm pt}}$. 
For instance, diagram Fig.~\ref{3ord_223B_1} is the result of the term
\ref{g4pt_2nd}($e+f$) and its exchange, on the right hand side of Fig.~\ref{g4pt_2nd}. 
Its conjugate self-energy diagram,~\ref{3ord_322B_1}, however,
is generated by the second contribution to $\Gamma^{6-{\rm pt}}$, Fig.~\ref{g6pt_2nd}b. 
This term is also related to diagram \ref{3ord}g. 
More specifically, the term~\ref{g6pt_2nd}b has 9 cyclic permutations of its 
indices, of which 6 contribute to diagram~\ref{3ord_322B_1} and 3 to diagram~\ref{3ord_322B_2}.
On the other hand, the conjugate of \ref{3ord_322B_2} is diagram~\ref{3ord_223B_2}, 
which is entirely due to the exchange contribution of the~\ref{g4pt_2nd}d term in $\Gamma^{4-{\rm pt}}$.
The direct contribution of this same term leads to diagram~\ref{3ord}c, which is already self-conjugate.

More importantly, however, nonperturbative self-energy expansions can be 
obtained by means of 
other hierarchy truncations at the level of $\Gamma^{4-{\rm pt}}$ and~$\Gamma^{6-{\rm pt}}$.
Translating these into self-energy expansions is then just an issue of
introducing them in Eq.~(\ref{irrself}).  
According to the approximation  chosen for the vertex functions appearing 
in Fig.~\ref{self_en}, we will be summing specific sets of diagrams when 
solving the Dyson equation, Eq.~(\ref{Dyson}). 
However, from the above discussion it should be clear that extra care must be taken
to guarantee that the truncations lead to physically coherent results. 
In particular, it is not always possible to naively neglect~$\Gamma^{6-{\rm pt}}$. 
The last two terms of the self-energy equation, Eq.~(\ref{irrself}), generate conjugate
contributions. Hence, neglecting one term or the other will spoil the 
analytic properties of the self-energy which require a Hermitian real part and
an anti-Hermitian imaginary part.
In the examples discussed above, diagrams \ref{3ord}f and \ref{3ord}g would
be missing if $\Gamma^{6-{\rm pt}}$ had not been considered. 

When no irreducible 3B interaction terms are present in the hierarchy truncation, 
only the $\Gamma^{4-{\rm pt}}$ term contributes to Eq.~(\ref{irrself}).
The ladder and the ring truncations, shown in
Eqs.~(\ref{GmLadd}) and (\ref{GmRing}) generate their own conjugate diagrams
and can be used on their own to obtain physical approximations to the self-energy. 
However, this need not be true in general. A counterexample is actually provided 
by the truncation of Eq.~(\ref{DeltaGm}) which, if inserted in Eq.~(\ref{irrself}) without
the corresponding contributions to $\Gamma^{6-{\rm pt}}$, 
cannot generate a correct self-energy.   Because of its diagrammatic content,
Eq.~(\ref{DeltaGm}) can only be used as a correction to~$\Gamma^{4-{\rm pt}}$.

\begin{figure}[t!]
\begin{center}
\includegraphics[scale=0.5]{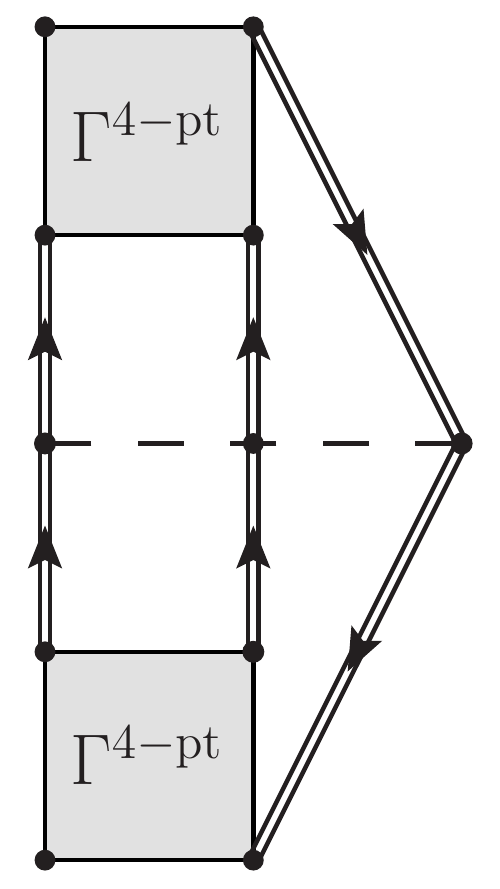}
\caption{Diagrammatic representation of the self-energy correction $\Delta\Sigma^{\star_{\Gamma W \Gamma}}$
 given in Eq.~(\ref{DeltaSig_G3G}).}
\label{fig:DeltaSig_G3G}
\end{center}
\end{figure}

As far as 3B interaction-irreducible diagrams are concerned, the most important 
contribution should be that associated with Fig.~\ref{3ord}c as discussed in Sec.~\ref{sec_PT_3ord}.
Further contributions with similar structures are also expected to contribute
to the correlation dynamics. To include such terms, 
one can go beyond third order by replacing the effective 2BFs
at the upper and lower ends by ladder or ring summations. 
Note that this is precisely the structure that arises from the hierarchy truncation associated 
to Eq.~(\ref{DeltaGm}). 
This would lead to a generalized contribution, whose diagrammatic content
is summarized by Fig.~\ref{fig:DeltaSig_G3G}. 
The corresponding expression for the self-energy would read:
\begin{widetext}
\beqn
\label{DeltaSig_G3G}
\Delta\Sigma^{\star_{\Gamma W \Gamma}}_{\al\be}(\om)=&& - \frac{(\ii\hbar)^4}{4}
\int\frac{\d\om_1}{2\pi}  \cdots \int\frac{\d\om_4}{2\pi}
\sum_{\substack{\ga\de\nu \\ \sig\tau\chi}} 
\sum_{\substack{\mu\ep\lm \\ \xi\eta\ta}}
\Gamma^{4-{\rm pt}}_{\al\ga,\de\nu}(\om,\om_1+\om_2-\om;\om_1,\om_2) 
G_{\de\mu}(\om_1)G_{\nu\ep}(\om_2) \,  \times
\\\nn &&
\qquad \qquad  
G_{\ta\ga}(\om_1+\om_2-\om) \, 
W_{\mu\ep\lm,\xi\eta\ta} \, G_{\xi\sig}(\om_3) G_{\eta\tau}(\om_4) \,
G_{\chi\lm}(\om_3+\om_4-\om) \, 
\Gamma^{4-{\rm pt}}_{\sig\tau,\be\chi}(\om_3,\om_4;\om,\om_3+\om_4-\om) \, .
\enqn
\end{widetext}
To quantify the importance of these terms, they would need to be included in the 
self-consistent procedure. Moreover, these corrections should also 
be considered when computing the total energy, as we will see next.
 
To conclude this Section, we would like to stress the fact that 
extensions to include 3BFs beyond effective 2B interactions, like $\widetilde{V}$,
are a completely virgin territory. To our knowledge, these have not been evaluated for
nuclear systems (or any other system, for that matter) with diagrammatic formalisms. 
Truncation schemes, like those proposed here, should provide
insight on in-medium 3B correlations. The advantage that the SCGF formalism
provides is the access to nonperturbative, conserving approximations 
that contain pure 3B dynamics without the need for \emph{ad hoc} 
assumptions. 
  
\section{Ground state energy}
\label{section4}

SCGF calculations aim at providing reliable calculations for the SP propagator of correlated
systems via diagrammatic techniques. Traditionally, there have been two motivations to do this.
On the one hand, the SP propagator provides access to all 1B operators and hence 
is a useful tool to characterize a wide range of the system's properties. On the other hand,
the ground state energy is a critically important 2B observable that can be obtained from the 1B GF 
itself. This 
is a crucial result, that arises from the GMK sum rule \cite{Gal58,Kol74}. The sum rule is valid both
at zero and at finite temperature, where the 1B propagator also provides access to the energy and, 
at least approximately, to all other thermodynamical properties \cite{Rios06}. 
In this Section, we investigate the modifications of the GMK sum rule when 3BF are
included in the Hamiltonian. 

Not all the information content from the propagator is needed to obtain the ground state energy. 
The hole part, which includes details
about the transition amplitude for the removal of a particle from the many-body system, is enough for 
these purposes. One therefore defines the diagonal part of the hole spectral function:
\beqn
\label{S_hole}
\mathcal{S}_\alpha^h(\om) &=& \frac{1}{ \pi} \mathrm{Im}\, G_{\al\al}(\om) 
\\\nn 
&=& \sum_n \left| \left\langle \Psi_n^{N-1} | a_\al | \Psi_0^N \right\rangle \right|^2 \delta \left[ \hbar \om - \left( E_0^N - E_n^{N-1} \right) \right] \, ,
\enqn
for energies below the Fermi energy, $\hbar \om < \varepsilon_F^- = E_0^N - E_0^{N-1}$. The 
$n^\text{th}$ excited state of the $N-1$ particle system is described by the many-body wave function
$| \Psi_n^{N-1} \rangle$ and has a total energy $E_n^{N-1}$. 
The transition amplitude between the $N$ and 
the $N-1$ body systems is closely related to the definition of the theoretical spectroscopic factor 
\cite{Dick04} and also provides information on removal strength
distributions. The complete spectral function
represents a direct link between theory and experiment,
as well as determining energy centroids \cite{Duguet12}.

To obtain the GMK sum rule, one starts by considering the first moment of the hole spectral function:
\beqn
\label{int_gmk}
I_\al &=& \int^{\epsilon^-_F}_{-\infty}\mathrm{d} \om \, \om\, \mathcal{S}_\alpha^h(\om)\,.
\enqn
From the spectral representation above, it is easy to see that this sum-rule is also the expectation
value over the many-body state of the  commutator:
\beqn
I_\al = \langle\Psi_0^N| \ad\al[\a\al,\h H]|\Psi_0^N\rangle\,.
\enqn
Using the Hamiltonian in Eq.~(\ref{H}), one can evaluate the commutator to find:
\beqn
\label{I_alpha}
I_\al=\langle\Psi_0^N|&&\sum_{\be} T_{\al\be}\, \ad\al \ad\be
+
\f 1 2\sum_{\ga\be\de} V_{\al\ga,\be\de}\, \ad\al\ad\ga a_\de a_\be
\\\nn
&&+ \frac{1}{12} \sum_{\substack{\ga\ep\be\de\eta}} W_{\al\ga\ep,\be\de\eta}\, 
\ad\al\ad\ga\ad\ep a_\eta a_\de a_\be |\Psi_0^N\rangle\,.
\enqn
Note that, in general, $T$ represents the 1B part of the Hamiltonian which, in addition to the kinetic
energy, might also contain the 1B potential. 
Summing over all the external SP states, $\alpha$, one finds,
\beq
\sum_\al I_\al=
\langle\Psi_0^N | \h T + 2 \h V + 3 \h W |\Psi_0^N\rangle\,.
\label{sumrule}
\enq
In other words, the sum over SP states of the first moment of the spectral function yields a particular
linear combination of the contributions  of the 1B, 2B and 3B potentials to the 
ground state energy,
\beq
E_0^N = 
\langle\Psi_0^N| \h H | \Psi_0^N\rangle 
=
\langle\Psi_0^N | \h T + \h V + \h W |\Psi_0^N\rangle\,.\label{gsenergy}
\enq

Since $\h T$ is a 1B operator, one can actually compute its expectation value from the SP
propagator itself:
\beq
\langle\Psi_0^N|\h T|\Psi_0^N\rangle
 =\frac{1}{\pi}\int^{\epsilon^-_F}_{-\infty} \mathrm{d} \om \sum_{\al\be} T_{\al\be} \mathrm{Im}\, G_{\be\al}(\om) \, .
\enq
The energy integral on the right hand side yields the 1B density matrix element, Eq.~(\ref{1B_densitymatrix}):
\beq
\rho^{1B}_{\be \al}=\frac{1}{\pi}\int^{\epsilon^-_F}_{-\infty} \mathrm{d} \om \, \mathrm{Im}\, G_{\be\al}(\om) \, ,
\enq
which can be used to simplify the previous expression. 
For the 2B case, this is enough to provide an independent constraint and hence allows for the calculation
of the total energy. The ground state energy can then be computed from the 1B propagator alone. 

When 3BF are present, however, one needs a third independent linear combination 
of $\langle \h T\rangle$, $\langle \h V\rangle$ and $\langle \h W\rangle$.
Knowledge of the 1B propagator is therefore not enough to compute the total energy, since either
the 2B or the 3B propagators are needed to compute $\langle \h V\rangle$ or $\langle \h W\rangle$ exactly.
Depending on which of the two operators is chosen, one is left with different expressions for the energy
of the ground state. This freedom in choice could in principle be exploited to test the validity of 
different approximations. In practical applications, however, one should choose the combination that provides minimum
uncertainty. 

Let us start by considering the case where the 3B operator is eliminated. Adding $2 \langle \h T\rangle$
and $\langle \h V\rangle$ to the sum rule, Eq.~(\ref{sumrule}), one finds the following exact expression for the total
ground state energy:
\beqn
\label{gmk_2b}
E_0^N&=&\frac{1}{3\pi}\int^{\epsilon^-_F}_{-\infty} \mathrm{d} \om \; \sum_{\al \be}(2T_{\al\be}+
\om \de_{\al\be}) \mathrm{Im}\, G_{\be\al}(\om) 
\nn \\
&&+\f 1 3 \langle\Psi_0^N|\widehat{V}|\Psi_0^N\rangle \, .
\enqn
The calculation of this expression requires the hole part of the 1B propagator and the two-hole
part of the 2B propagator, which would appear in the second term. 
We note that this expression is somewhat equivalent to the original GMK, in that the ground state energy
is computed from 1B and 2B operators, even though the Hamiltonian itself is a 3B operator. 
This might prove advantageous in calculations where the 2B propagator is computed explicitly. 

Alternatively, one can eliminate the 2B contribution from the GMK sum rule by adding $\langle \h T\rangle$
and subtracting $\langle \h W\rangle$ to the sum rule, Eq.~(\ref{sumrule}). This leads to the expression:
\beqn
\label{gmk_3b}
E_0^N&=&\frac{1}{2\pi}\int^{\epsilon^-_F}_{-\infty} \d\om \; \sum_{\al\be}(T_{\al\be}+
\om \de_{\al\be}) \mathrm{Im}\, G_{\be\al}(\om) \, 
\nn \\
&& - \f 1 2 \langle\Psi_0^N| \h W |\Psi_0^N\rangle
\enqn
The first term in this expression is formally the same as that obtained in the case where only 2BFs are
present in the Hamiltonian. In that sense, the second term can be thought of as a correction to the 
total energy associated with the 3BF. Note, however, that the 3BF does influence the 1B propagator on the first term
and hence the correction should only be applied at the very end of the self-consistent procedure. 

Eqs.~(\ref{gmk_2b}) and~(\ref{gmk_3b}) are both exact. 
Which of the two is employed in actual calculations will mostly depend on the accuracy 
associated with the evaluation of the expectation values, 
$\langle \h V \rangle$ and $\langle \h W \rangle$. 
If the 2B interaction is dominant with respect to the 3BF, for instance, the former will
be a large contribution. Small errors in the calculation of the 2B propagator could eventually yield
artificially large corrections in the ground state energy. In nuclear physics, the 3BF expectation value
is expected to provide a smaller contribution than the 2BF~\cite{grange1989,Epel09}. 
Consequently, approximations in Eq.~(\ref{gmk_3b}) should lead to smaller absolute errors.
This has been the approach that we have recently followed
in both finite nuclei and infinite nuclear matter~\cite{Cipol13,Car13}. 
In finite nuclei, evaluating $\langle \h W \rangle$ at first order in terms of \emph{dressed} propagators
leads to satisfactory results. However, accuracy is lost if free propagators, $G^{(0)}$ are used instead.
Eq.~(\ref{gmk_2b}) may eventually be useful in calculations of infinite matter, in which the $\Gamma^{4-{\rm pt}}$ is
calculated nonperturbatively.

This first-order approximation with undressed propagators is traditionally used in nuclear structure.
In this context, three-body forces have been often discussed in the Hartree-Fock 
approximation with Skyrme or Gogny functionals \cite{Ring84,Bender03}. Zero-range forces have also
been employed in \emph{ab-initio}-type calculations \cite{Gunther10,*vanDalen09}. It is perhaps instructive
to point out at this stage that the previous formulae apply to this case as well. In particular,
the Hartree-Fock approximation with 3BF can be alternatively derived from the variational principle,
under the assumption that the many-body state is described by a Slater determinant, $\rphizero$. 
Diagonalizing an effective 1B hamiltonian leads to a series of Hartree-Fock orbitals with single-particle energies,
$\varepsilon_\alpha$. The total energy under a 2B Hamiltonian is not the sum of these energies, but
rather requires a correction to avoid double-counting \cite{Ring84}. Similarly, in the 3B case,
the energy is computed as follows:
\beqn
\label{gmk_HF}
E_0^{HF}&=& \sum_\alpha \varepsilon_\alpha 
- \langle \h V \rangle_{HF}
- 2 \langle \h W \rangle_{HF} \, .
\enqn
This result is straightforwardly derived by noticing that, in the Hartree-Fock approximation, the sum rule, 
Eq.~(\ref{sumrule}), reduces to the first term. Within this approximation, the expectation values 
can be directly computed from the uncorrelated 1B density matrix:
\beqn
\langle \h V \rangle_{HF} &=& \frac{1}{2} \sum_{\substack{\al \ga \\ \be \de} } V_{\al\ga,\be\de} 
\rho^{HF}_{\be\al} \rho^{HF}_{\de\ga} \, ,\\
\langle \h W \rangle_{HF} &=&  \frac{1}{6} \sum_{\substack{\al \ga \ep \\ \be \de \eta }} W_{\al\ga\ep,\be\de\eta} 
\rho^{HF}_{ \be\al} \rho^{HF}_{\de\ga} \rho^{HF}_{\eta\ep} \, .
\label{HF3BF}
\enqn
If the 3B contribution is perturbative, one would expect Eq.~(\ref{HF3BF}) to provide 
a good starting point to evaluate the full 3BF contribution of Eq.~(\ref{gmk_3b}). This seems to be the case in
finite nuclei where, however, the internal density matrices should be appropriately dressed \cite{Cipol13} 
to find accurate results.

To close this section, let us comment on the use of effective interactions in the calculation of
the ground state energy itself. Two errors can arise in this context. The first would involve incorrectly
accounting for the different pre-factor in the 1B and 2B effective interactions. 
This double counting of the HF potential has already been discussed at the end of Sec.~\ref{section2}.
The second issue would arise if a 3B correction to the energy was neglected.
The Hartree-Fock case provides a good example of the latter. Replacing
the bare 2B interaction in Eq.~(\ref{gmk_HF}) by the effective 2B force would immediately lead to errors. 
The 3B correction in Eq.~(\ref{gmk_HF}) would necessarily have to change to provide the same result.
Consequently, performing a calculation with an effective 2B force and simply computing the energy
with the usual 2B formulae is not enough. The 3B correction is needed anyway and is different if one uses the bare or the effective interaction. 

\section{Conclusions}
\label{concl}

We have presented an extension of the SCGF approach to include 3B interactions. 
The method allows to incorporate consistently 2B and 3B forces on an equal footing and 
should be interesting for nuclear physics applications. Other many-body systems in which
induced 3BF are generated by cuts in the model space could potentially benefit from this treatment as well.

The 3BF has been introduced in two different but equivalent ways in the formalism. 
On the one hand, we have studied the diagrammatic perturbative expansion of the propagator
including 1B, 2B and 3B forces. The expansion is analogous to cases previously studied in the literature,
but the 3BF requires some careful handling. We present in Appendix~\ref{app_Feyn} the Feynman
rules associated with this expansion. 
Within a SCGF approach, where propagators are dressed and the Dyson equation is used iteratively,
only 1PI skeleton diagrams enter the expansion. The number of diagrams can be further
reduced by introducing effective interactions, which sum up sub-series of interaction-reducible
diagrams. These effective interactions can be interpreted as a generalization of the normal ordering 
of the many-body hamiltonian. Instead of ordering with respect to an uncorrelated state, however, the
effective interactions include the effect of many-body correlations by construction. 
The proper self-energy can be defined from 1B, 2B and 3B forces and still be computed within the Dyson's equation. 
We have shown how this effective grouping of operators reduces the number of diagrams by considering
the perturbation expansion up to third order. The equivalence between the original diagrammatic expansion
and that obtained from the effective interaction at any arbitrary order is proven in Appendix~\ref{app_Heff}.

On the other hand, the propagator method can be expressed using the EOM. We have re-derived the
basic equations of this method, consistently including 3BFs. The Martin-Schwinger hierarchy now connects
the $(n)$-body propagator to the $(n-1)$-, the $(n+1)$- and, via the 3BF, the $(n+2)$-body GF. 
In turn, this requires the introduction of vertex functions beyond the 4-point level. 
Through the hierarchy of the EOM, we have found an expression for the vertex function $\Gamma^\text{4-{\rm pt}}$,
which embodies the higher order interacting contributions beyond the mean-field. Truncation to second order
of this function, together with complete second order expression for the 6-point $\Gamma$ function, provides the third order
approximation for the irreducible self-energy. The correspondence to the diagrams obtained in the perturbative
expansion indicates that these two different approaches are equivalent.

Moreover, we have shown that, using the 2B effective interaction in truncation 
schemes based on $\Gamma^\text{4-{\rm pt}}$, leads to either ladder, ring or parquet 
approximations that effectively
include some 3B terms. Within these approximations, the general structure of the formalism, based on 2BFs, remains
unaltered \cite{Dick04}. Results obtained recently both in calculations for infinite nuclear matter
\cite{Car13} as well as nuclei \cite{Cipol13} exploit this expanded self-consistent Green's functions approach to include 3B nuclear forces.

More importantly, however, this approach is able to provide a general mechanism
to devise nonperturbative resummation schemes. In particular, we have proposed a potentially relevant 
correction to the self-energy that includes interaction-irreducible three-body effects explicitly. 
Extensions to include 3BFs beyond the effective 2BF approach are lacking in the literature and could prove
to be potentially relevant in some instances, particularly in nuclear physics. 

Finally, we have presented a general method to compute the energy of the many-body ground state by 
means of the GMK sum rule. The sum rule still allows for the calculation of the ground state energy from the
$2(n-1)$-point GFs in spite of the fact that the energy itself is an $n$-body operator. 
Two possible approaches have been proposed, which require the calculation of either a 2B or a 3B 
expectation value. Depending on the relative importance of the 2B and the 3B forces in the system, one or 
the other might be preferable. 

Calculations performed using this extended SCGF formalism have already been presented in the literature \cite{Cipol13,Car13}. 
This expanded approach provides a firm basis for further studies of nuclear systems from a Green's functions point of 
view. The formalism can be extended to finite temperature and off-equilibrium settings. More importantly, it provides
a framework to compute many  relevant quantities for the description of a quantum many-body system,
from binding energies, to thermodynamical properties or even pairing.
On the same footing, the Gorkov-Green's function formalism for finite nuclei could be 
improved to include 3BFs.

We believe that this extended approach is an interesting tool to study quantum many-body systems 
from an \emph{ab-initio} microscopic point of view. In principle, the framework provides a coherent description of 
the correlated, nonperturbative dynamics of systems with many-body interactions. The generalization 
to Hamiltonians including $N$-body forces can be performed along similar lines. In addition to its academic interest,
advances in interaction-tunable ultra cold gases might require these developments. In nuclear physics, the
importance of 4BFs, either bare or induced, could also be assessed using analogous techniques. Ultimately, 
the methods presented here should be a good starting point to foster new initiatives that systematically 
address the issue of many-body forces in quantum many-body systems.

\begin{acknowledgments}
This work is supported by the Consolider Ingenio 2010 Programme CPAN CSD2007-00042, 
Grant No. FIS2011-24154 from MICINN (Spain) and 
Grant No.~2009SGR-1289 from Generalitat de Catalunya  (Spain);
and by STFC, through Grants ST/I005528/1 and ST/J000051/1. 
A. Carbone acknowledges the support by the SUR of the ECO from Generalitat de Catalunya.
\end{acknowledgments}

\appendix
\section{Feynman diagram rules for 2- and 3-body interactions}
\label{app_Feyn} 

We present in this appendix the Feynman rules associated with the diagrams arising
in the perturbative expansion of Eq.~(\ref{gpert}). The rules are given both in time and energy formulation, and some specific examples will be considered at the end. We pay 
particular attention to non-trivial symmetry factors arising in diagrams that include many-body 
interactions.
We work with antisymmetrized matrix elements, but for practical purposes represent 
them by extended lines. 

We provide the Feynman diagram rules for a given $p$-body propagator, such as Eqs.~(\ref{g4pt}) 
and~(\ref{g6pt}). These arise from a trivial generalization of the perturbative 
expansion of the 1B propagator in Eq.~(\ref{gpert}). 
At  $k$-th order in perturbation theory, any contribution from the time-ordered product in 
Eq.~(\ref{gpert}), or its generalization, is represented by a diagram with $2p$ external 
lines and $k$ interaction lines (from here on called vertices), 
all connected by means of oriented fermion lines. 
These fermion lines arise from contractions between annihilation and creation operators,
$$
\,\,\underbracket[0.6pt][0.2em]{\!\!a^I_{\de}(t)a\!\!}\,{}^{I \, \dag}_{\gamma}(t') \equiv \lan \Phi_0^N | \T \left[ a^I_{\de}(t)a^{I \, \dag}_{\gamma}(t')\right] |\Phi_0^N\ran=\ii\hbar \,G^{(0)}_{\de\gamma}(t-t') .
$$ 
Applying the Wick theorem to any such arbitrary diagram, results in the following Feynman rules.
\begin{description}
\item[Rule 1] Draw all, topologically distinct and connected diagrams with $k$ vertices, and $p$ incoming and $p$ outgoing external lines, using directed arrows. For interaction vertices the external lines are not present.
\item[Rule 2] Each oriented fermion line represents a Wick contraction, leading to the unperturbed propagator  
$\ii\hbar G_{\al\be}^{(0)}(t-t')$ [or $\ii\hbar G_{\al\be}^{(0)}(\omega_i)$]. 
In time formulation, the $t$ and $t'$ label the times of the vertices at the end and at the beginning of the line. 
In energy formulation, $\omega_i$ denotes the energy carried by the propagator. 
\item[Rule 3] Each fermion line starting from and ending at the \emph{same} vertex is an 
equal-time propagator,  $-\ii\hbar G_{\al\be}^{(0)}(0^-)=\rho_{\al\be}^{(0)}$.
\item[Rule 4] For each 1B, 2B or 3B vertex, write down a factor $\frac{\ii}{\hbar} U_{\al \be}$, \, $-\frac{\ii}{\hbar} V_{\al\ga,\be\de}$  or  $-\frac{\ii}{\hbar} W_{\al\ga\xi,\be\de\ta}$, respectively. For effective interactions, the factors are $-\frac{\ii}{\hbar} \widetilde{U}_{\al \be}$, \, $-\frac{\ii}{\hbar} \widetilde{V}_{\al\ga,\be\de}$.
\end{description}
When propagator renormalization is considered, only skeleton diagrams are used in the 
expansion. In that case, the previous rules apply with the substitution 
$\ii \hbar G_{\alpha\beta}^{(0)} \to 
\ii \hbar G_{\alpha\beta}$.
Furthermore, note that  Rule 3 applies to diagrams embedded 
in the one-body effective interaction 
(see Fig.~\ref{ueffective}) and therefore they should not be considered explicitly in
an interaction-irreducible expansion. 
In calculating $\tilde U$, however, 
one should use the correlated $\rho_{\alpha\beta}$ instead of the unperturbed one. 
\begin{description}
\item[Rule 5] Include a factor $(-1)^{L}$ where $L$ is the number of closed fermion loops. This sign comes from the odd permutation of  operators needed to create a loop
and does not include loops of a single propagator, already accounted for by Rule 3.
\item[Rule 6] For a diagram representing a $2p$-point GF, add a factor $(-\ii/\hbar)$, whereas for a $2p$-point interaction vertex without external lines (such as $\Sigma^\star$ and $\Gamma^{2p-pt}$) add a factor $\ii\hbar$.
\end{description}
The next two rules require a distinction between the time and the energy representation. 
In the time representation:
\begin{description}
\item[Rule 7] Assign a time to each interaction vertex. All the fermion lines connected to the same vertex $i$ share the same time,~$t_i$. 
\item[Rule 8] Sum over all the internal quantum numbers and integrate over all internal times from $-\infty$ to $+\infty$. 
\end{description}
Alternatively, in energy representation:
\begin{description}
\item[Rule 7']  Label each fermion line with an energy $\omega_i$, 
under the \emph{constraint} that the total incoming energy equals the total outgoing energy at 
each interaction vertex, \hbox{$\sum_i\omega_i^{in}=\sum_i\omega_i^{out}$}.
\item[Rule 8'] Sum over all the internal quantum numbers and integrate over each independent internal energy, with an extra factor $\frac{1}{2\pi}$, i.e. $\int^{+\infty}_{-\infty} \frac{d\omega_i}{2\pi}$.
\end{description}

\begin{figure}[t!]
  \subfloat[]{\label{diagr_29}\includegraphics[scale=0.6]{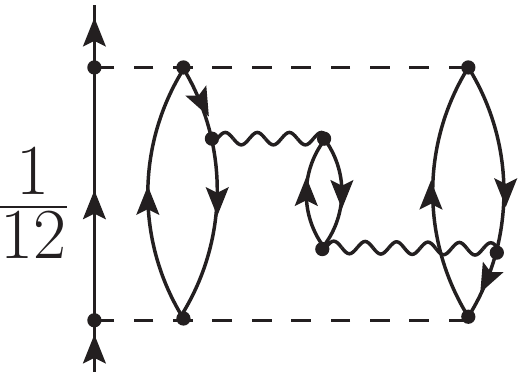}}
  \hspace{1cm}
  \subfloat[]{\label{diagr_27}\includegraphics[scale=0.6]{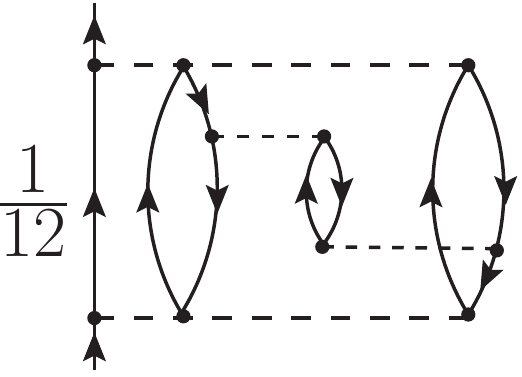}}
    
  \subfloat[]{\label{diagr_26}\includegraphics[scale=0.6]{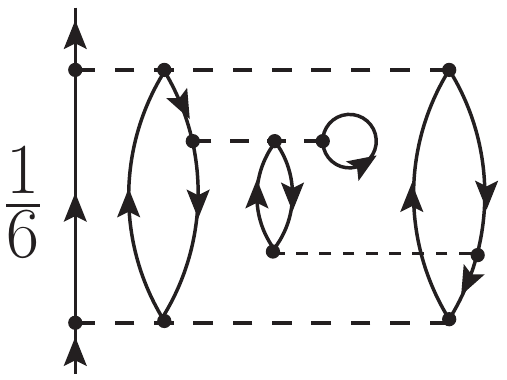}}
   \hspace{0.8cm}
  \subfloat[]{\label{diagr_28}\includegraphics[scale=0.6]{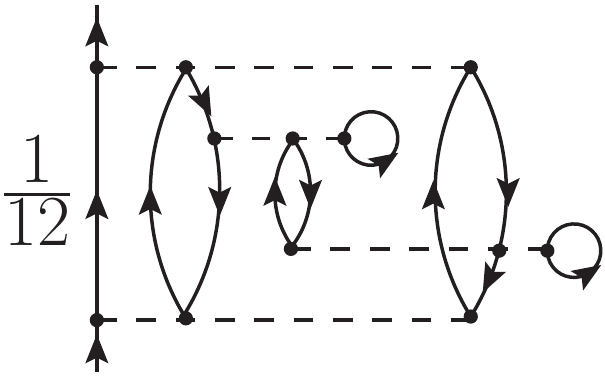}}
\caption{Examples of diagrams containing symmetric and interacting lines, with explicit symmetry 
factors. Diagrams (b) to (d) are obtained by expanding the effective
interaction of diagram (a) according to Eq (\ref{veff}). 
Swapping the 3B and 2B internal vertices in (c) gives a distinct, but topologically equivalent, contribution.}
\label{rule9-2a}
\end{figure}

Each diagram is then multiplied by a combinatorial factor S that  originates from the number of 
equivalent Wick contractions that lead to it. This symmetry factor 
represents the order of the symmetry group for one specific diagram or, in other words, 
the order of the permutation group of both open and closed lines, once the vertices are fixed. 
Its structure, assuming only 2BFs and 3BFs, is the following :
\beq
S=\frac{1}{k!}\frac{1} {[(2!)^2]^{q} [(3!)^2]^{k-q} }\binom{k}{q} \; C
= \prod_i S_i \; .
\label{diagsymfac}
\enq
Here, $k$ represents the order of expansion. 
$q$ ($k-q$) denotes the number of 2B (3B) vertices in the diagram.
The binomial factor counts the number of terms in the expansion $(V+W)^k$ 
that have $q$ factors of $V$ and $k-q$ factors of $W$.
Finally, $C$ is  the overall number of \emph{distinct} contractions and reflects 
the symmetries of the diagram. Stating general rules to find $C$ is not simple. 
For example, explicit simple rules valid for the well-known $\lambda \phi^4$ scalar  theory are still 
an object of debate~\cite{Feyn_rules}. 
An explicit calculation for $C$ has to be carried out diagram by diagram 
\cite{Feyn_rules}. Eq.~(\ref{diagsymfac}) can normally be factorized in a product factors $S_i$,
each due to a particular symmetry of the diagram. In the following, we list a series of specific examples which is,
by all  means, not exhaustive.
\begin{description}
 \item[Rule 9]  For each group of $n$ symmetric lines, or symmetric groups-of-lines as defined below, multiply by a symmetry factor $S_i$=$\frac{1}{n!}$. The overall symmetry factor of the diagram will be $S=\prod_i S_i$.
Possible cases include:
  \end{description}
\begin{enumerate}[(i)]
 \item {\em Equivalent lines}.  
 $n$ equally-oriented fermion lines are said to be equivalent if they start from the same initial vertex and end on the same final vertex.
 \item {\em Symmetric and interacting lines}.  
 $n$ equally-oriented fermion lines that start from the same initial vertex and end on the same final 
 vertex, but are linked via an interaction vertex to one or more close fermion line blocks. 
 The factor arises as long as the diagram is {\em invariant} under the permutation of the two blocks.
 \item {\em Equivalent groups of lines}. 
 These are blocks of interacting lines (e.g. series of bubbles) that are equal to each other: 
           they all start from the same initial vertex and end on the same final vertex.
 \end{enumerate} 

 Rule 9-i  is the most well-known case and applies, for instance, to the two second order diagrams 
 of Fig.~\ref{2ord}. Diagram \ref{2ord}a has 2 upward-going equivalent lines and requires a symmetry factor $S_e$=$\frac1{2!}$. In contrast,Êdiagram \ref{2ord}b has 3 upward-going equivalent lines and 2 downward-going equivalent lines, that give a factor $S_e$=$\frac1{2! \, 3!}$=$\frac1{12}$.
 
\begin{figure}[t!]
  \subfloat[]{\label{diagr_33}\includegraphics[scale=0.65]{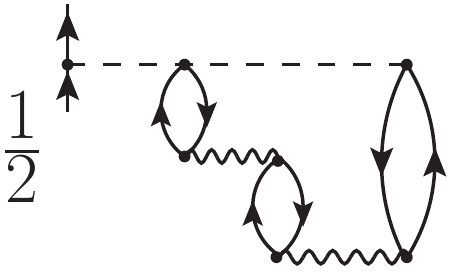}}
  \hspace{1cm}
  \subfloat[]{\label{diagr_31}\includegraphics[scale=0.65]{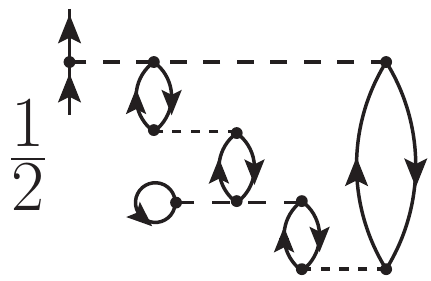}}
    \newline
  \subfloat[]{\label{diagr_12}\includegraphics[scale=0.65]{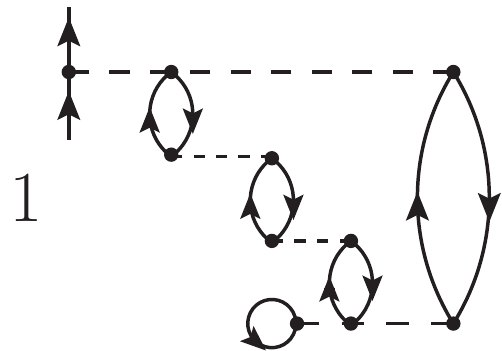}}
\caption{Examples of diagrams entering the static part of the self-energy. Applying rule 9-ii, 
diagrams (a) and (b) take a factor $S_{si}=\frac{1}{2}$ from the symmetry between the 
two bubbles attached to the upper three body vertex. 
The symmetry is broken in diagram (c), where the overall factor is  $S_{si}=1$  }
\label{rule9-2b}
\end{figure}

\begin{figure*}[th!]
 \includegraphics[scale=0.8]{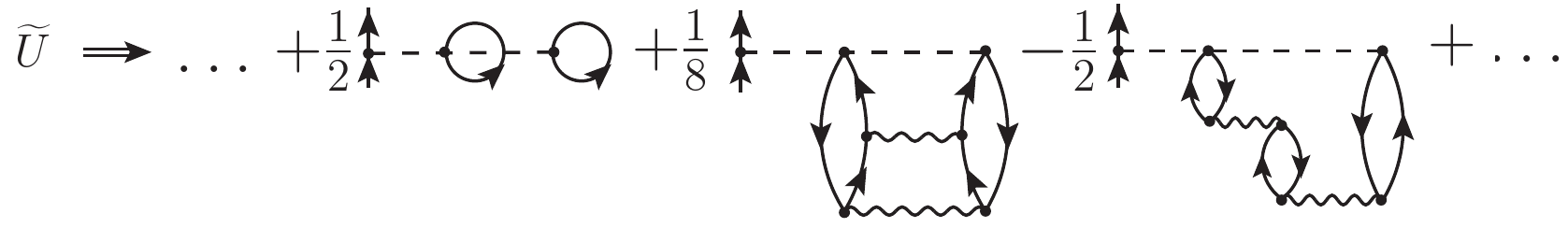}
\caption{Diagrams entering the effective one-body interaction, Eq.~(\ref{ueff}), obtained by 
substituting the right hand side of Fig.~\ref{g4pt_2nd} into Eq.~(\ref{g4ptgamma}). The two bubble terms 
correctly reproduce the symmetric factor inferred by applying rules 9-i and 9-ii. }
\label{rule9-2u}
\end{figure*}

Figs.~\ref{rule9-2a} and~\ref{rule9-2b} give specific examples of the application of rule 9-ii. 
Diagram~\ref{diagr_29} has 3 upward-going equivalent, non-interacting lines, which yield a
symmetry factor $S_e$=$\frac1{3!}$ due to rule 9-i. 
However, there are also two downward-going symmetric and equivalent lines, that 
interact through the exchange of a bubble and thus give rise to a factor $S_{si}$=$\frac1{2!}$. 
The total factor is therefore $S$=$S_{e} \times S_{si}$=$\frac1{12}$.
Let us now expand the two 2B effective interactions that are connected to the intermediate
bubble according to Eq.~(\ref{veff}).
Diagram~\ref{diagr_29}  is now seen to contain three contributions, diagrams~\ref{diagr_27} to 
\ref{diagr_28}, with the symmetry factors shown in the figure.
Note that drawing the contracted 3B vertex above or below the bubble 
in \ref{diagr_26} leads to two topologically equivalent diagrams that must only be drawn once, 
i.e. diagram~\ref{diagr_26}. However, since the diagram is no longer symmetric under the 
exchange of the 
two downward-going equivalent lines, rule 9-ii does not apply anymore and the $S_{si}$
factor is no longer needed. 

A similar situation occurs when the two interacting fermion lines start and end on the same vertex, as in Fig.~\ref{rule9-2b}. Consider the left-most and right-most external fermion bubbles. 
In all three diagrams, they are
connected to each other by a 3B interaction vertex above and by a series of interactions and
medium polarizations below. 
The intermediate bubble interactions in diagrams \ref{diagr_33} and \ref{diagr_31} are symmetric under 
exchange. There are therefore two sets of symmetric interacting lines (the two up-going and two 
down-going fermion lines) and hence both diagrams take a factor $S_{si}=1/2$. 
In contrast, the two external loops in \ref{diagr_12} are not symmetric under exchange due to the
lower 3B vertex. Rule 9-ii does not apply anymore and  $S_{si}=1$. 
If all the vertices between the external loops where equal (e.g. effective 2B terms $\tilde{V}$),
a factor $S_{si}$=$1/2$ would still apply.

The case of Fig.~\ref{rule9-2b} is of particular importance because these diagrams directly contribute to 
the energy-independent 1B effective interaction. 
In the EOM approach, these contributions arise from 
the first three terms on the right hand side of Fig.~\ref{g4pt_2nd}. Note that the ladder diagram 
has a symmetry factor $S_{e}$=$1/2$ and that the exchange contribution in the bubble term has to be 
considered. Using these diagrams to the define the 2B propagator in Eq.~(\ref{g4ptgamma}) and 
inserting these in the last term of Eq.~(\ref{ueff}), one finds the contributions to $\tilde{U}$ shown 
in Fig.~\ref{rule9-2u}. The two bubble terms have summed up to form diagram \ref{rule9-2b}a, each 
of them contributing a factor $1/4$ from Eq.~(\ref{ueff}). Consequently, the approach leads to 
the correct overall $S_{si}$=$1/2$ symmetry factor.
In our approach, there is no need to explicitly
compute these diagrams, since they are automatically included by Eq.~(\ref{ueff}).

\begin{figure}[t]
  \centering
  \subfloat[]{\label{diagr_32}\includegraphics[scale=0.7]{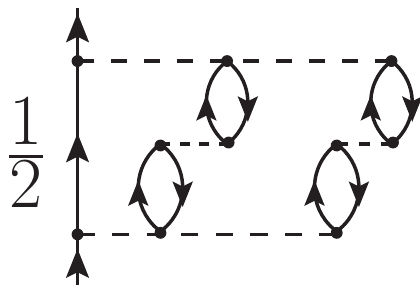}}
  \hspace{0.6cm}
  \subfloat[]{\label{diagr_30}\includegraphics[scale=0.6]{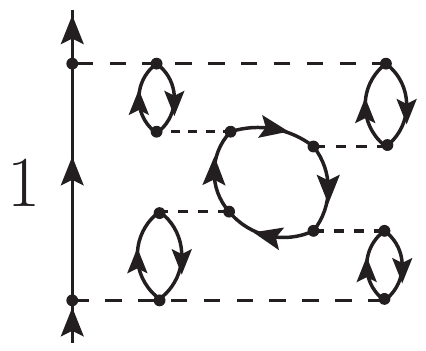}}
  \caption{ Examples of a diagram where equivalent group of lines are present and one where rule 9-iii does not apply. Swapping the two chains of bubbles in (a), one finds an identical diagram. This is precisely the case of rule 9-iii, which brings in a factor $S_{egl}$=$\frac{1}{2}$. Performing the same exchange in diagram (b) generates a graph where the direction of the internal loop is reversed. No symmetry rule applies here and $S_{egl}$=$1$}
\label{rule9-3}
\end{figure}

Finally, rule 9-iii applies to the diagram in Fig.~\ref{diagr_32}. In this case, the two chains of bubble 
diagrams are equal and start and end at the same 3BF vertices. Hence, they are equivalent groups of 
lines and the diagram takes a factor $S_{egl}=\frac{1}{2}$. 
Diagram~\ref{diagr_30} is different because the exchange of all the bubbles 
generates a diagram in which the direction of the internal fermion loop is reversed. 
Therefore no symmetry rule applies and the symmetry factor is just $S_{egl}=1$.
This is, however, topologically equivalent to the initial diagram and hence must be 
counted only once.

As an example of the application of the above Feynman rules, we give here the formulae for some of the diagrams in Fig.~\ref{3ord}.
Let us start by a contribution that has been discussed in Section~\ref{section3}, diagram~\ref{3ord_232B}. There are two sets of upward-going equivalent lines, which contribute to a
symmetry factor $S_e=\frac{1}{2^2}$. Considering the overall factor of Eq.~(\ref{diagsymfac}) and the
presence of one closed fermion loop, one finds:
\begin{widetext}
\beqn
\Sigma^{(5c)}_{\al \be}(\om)=
- \frac{(\ii \hbar)^{4} }{4}
\int\frac{\d\om_1}{2\pi} \cdots \int\frac{\d\om_4}{2\pi}
\sum_{\substack{ \ga\de\nu \mu\ep\lm \\ \xi\eta\ta \sig\tau\chi}} 
&&
\widetilde{V}_{\al\ga,\de\nu}\gz_{\de\mu}(\om_1)\gz_{\nu\ep}(\om_2) 
W_{\mu\ep\lm,\xi\eta\ta}\gz_{\xi\sig}(\om_3)\gz_{\eta\tau}(\om_4) \times 
\\\nn &&
\gz_{\ta\ga}(\om_1+\om_2-\om)
\widetilde{V}_{\sig\tau,\be\chi}
\gz_{\chi\lm}(\om_3+\om_4-\om)  \, .
\enqn
 \end{widetext}
Diagrams \ref{3ord_233B_1} and \ref{3ord_233B_2} differ only for the orientation of a loop. Hence, there are
two pairs of equivalent lines in the first case and one pair and one triplet of equivalent lines in the second, which is reflected in their
different symmetry factors:
\begin{widetext}
\beqn
\Sigma^{(5h)}_{\al \be}(\om)=\frac{(\ii \hbar)^{5} }{4}
\int\frac{\d\om_1}{2\pi} \cdots \int\frac{\d\om_5}{2\pi}
\sum_{\substack{ \ga\de\ep \\ \xi\ta\sig \mu\nu\lm \\ \eta\tau\phi \chi\zeta}} 
&&
\widetilde{V}_{\al\ga,\de\ep}  \gz_{\de\xi}(\om_1)  \gz_{\nu\ga}(\om_2) 
W_{\xi\ta\sig, \mu\nu\lm}  \gz_{\mu\eta}(\om_3)\gz_{\chi\ta}(\om_4)\times
\\ \nn &&
\gz_{\ep\tau}(\om-\om_1+\om_2) \, W_{\eta\tau\phi,\beta\chi\zeta } 
\, \gz_{\lm\phi}(\om_5) \, \gz_{\zeta \sig} (\om_2+\om_3+\om_5-\om_1-\om_4) 
\, ,
\enqn
\beqn
\Sigma^{(5i)}_{\al \be}(\om)=\frac{(\ii \hbar)^{5} }{12}
\int\frac{\d\om_1}{2\pi} \cdots \int\frac{\d\om_5}{2\pi}
\sum_{\substack{ \ga\de\ep \\ \xi\ta\sig \mu\nu\lm \\ \eta\tau\phi \chi\zeta}} 
&&
\widetilde{V}_{\al\ga,\de\ep}  \gz_{\de\xi}(\om_1)  \gz_{\ep\ta}(\om_2) 
W_{\xi\ta\sig, \mu\nu\lm}  \gz_{\mu\eta}(\om_3)\gz_{\nu\tau}(\om_4)\times
\\ \nn &&
\gz_{\chi\ga}(\om_1+\om_2-\om) \, W_{\eta\tau\phi,\beta\chi\zeta } 
\, \gz_{\lm\phi}(\om_5) \, \gz_{\zeta \sig} (\om_3+\om_4+\om_5-\om_1-\om_2) 
\, .
\enqn
 \end{widetext}

\section{Interaction-irreducible diagrams with effective 1B and 2B interactions at all orders}
\label{app_Heff}

Interaction-irreducible diagrams can be used to distinguish between two 
different many-body effects in
the SCGF approach. On the one hand, effective interactions 
sum all the instantaneous contributions associated with ``averaging out" subgroups
of particles that lead to interaction-reducible diagrams. This has the advantage of
reducing drastically the number of diagrams at each order in the perturbative expansion.
It also gives rise to well-defined in-medium interactions. 
On the other hand, the remaining diagrams will now include higher-order terms summed
via the effective interaction itself. 

In this Appendix, we proove that the perturbative expansion can be recast 
into a set containing only interaction-irreducible diagrams at any given order, as long as
properly defined effective interactions are used. 
The argument we propose has been often used to demonstrate how disconnected 
diagrams cancel out in the perturbative expansion. 
We now apply it to a slightly different case that requires extra care. 
We focus on the case of a diagram that includes only 2B and 3BFs. 
The extension to the general case of many-body forces should be straightforward.

Eq. (\ref{gpert}) gives the perturbative expansion of the 1B GF in terms of the Hamiltonian, $H(t)$, in the interaction picture. The $k$-th order term of the perturbative expansion reads:
\begin{widetext}
\begin{align}
\label{vterm}
G_{\al\be}^{(k-{\rm th})}(t-t')=\left(\frac{-\ii}{\hbar}\right)^{k+1}\frac{1}{k!}\underbracket{\idotsint \textrm{d} t_{k}}_{k \,\,\textrm{terms}}
\lan \Phi_0^N | \T \left[ a^I_{\al}(t) a_{\be}^{I \dg} (t')H(t_1)\cdots H(t_k)\right] |\Phi_0^N\ran_\text{conn} \; .
\end{align}
\end{widetext}
Only connected contributions are allowed and we take the interaction picture
 external creation and destruction operators to the left for convenience. Let us assume, without loss of generality, 
that the diagram is composed of $q$ 2B and $k-q$ 3B interaction operators. 
This gives rise to $\binom{k}{q}$ identical contributions when expanding
 \hbox{$H(t)=T(t)+V(t)+W(t)$} in the time-ordered product, as discussed right after Eq.~(\ref{diagsymfac}) above.
 
Let us denote with $O(t)$ a generic operator, representing either a 2B, 
$V(t)$, or a 3B, $W(t)$, potential. 
Suppose now that there is a sub-set of $m$ operators that are arbitrary connected to
each other, but that share the external links with a \emph{unique} operator, $O(t_n)$, outside 
the subset. In other words, $O(t_{n})$ is the only way to enter and exit the subset of $m$-
operators $\{O(t_{n+1}), \cdots, O(t_{n+m})\}$  as drawn below:
\begin{equation}
\label{red_cut}
O\!\!\!\overbracket{\,\,(t_1)\cdots O\!\!\!\underbracket{\,\,(t_{n-1})\,\cdot\,\,\, }}\!\!O\!\!\!\Bigg{\vert}\phantom{a}\!\!\!\overbracket{\underbracket{ \overbracket{\,(t_n)\cdot
\,\Big\{\,O \!\!}\,\,(t_{n+1})\cdots\!\!\!}\,\,\, O\!\!}\,\,(t_{n+m})\,\,\Big\}
\end{equation}
$O(t_n)$ is also necessarily connected to the other interactions and, hence, this is an 
\emph{articulation vertex}. 
In general, there can be an arbitrary number of articulation vertices, such as  $O(t_n)$, at any 
given order. Each one of these vertices would isolate a particular subset of operators. The following arguments can be applied to each subset separately. 

For simplicity, let us restrict the argument to the simplest case of one articulation vertex only. Suppose that, among $m$ terms, there are $a$ 2B and $b$ 3B interactions, with $a+b=m$.
The number of time-ordered products $V(t)$ and $W(t)$ in Eq.~(\ref{vterm}) that is consistent with the above decomposition is
\begin{equation}
\binom{k}{q}\binom{q}{a}\binom{k-q}{b}=\binom{k}{m}\binom{m}{a}\binom{n}{q-a}
\end{equation}
where $m+n=k$. 

Let us consider the case in which $O(t_n)$ is a 3B operator, with matrix elements 
$W_{\mu\ga\de,\ta\sig\xi}$ connected with four legs to the internal subset of $m$ vertices and 
with two legs to the rest of the diagram. We can factorize the
amplitude in Eq.~(\ref{vterm}) by adding an intermediate identity operator as follows:
\begin{widetext}
\begin{eqnarray}
\label{redu}
&&\frac{1}{n!}\binom{n}{q-a}\underbracket{\idotsint \textrm{d} t_{n}}_{n \,\,\textrm{terms}} 
\lan\Phi_0^N | \T \left[ a^I_{\al}(t) a^{I \dag}_{\be}(t')O(t_1) ~ \cdots ~ O(t_{n-1})\, a^{I \dag}_{\mu}(t_n^+)a^I_{\ta}(t_n)\right] |\Phi_0^N \ran \,\,W_{\mu\ga\de,\ta\sig\xi} \, \,\, \frac{1}{(3!)^2} \binom{3}{1}^2  
\nn \\
&& \quad \quad \times \quad
  \frac{1}{(m)!}\binom{m}{a}\underbracket{\idotsint  \textrm{d} t_{k}}_{m \,\textrm{terms} }  \lan \Phi_0^N | \T \left[a^{I\dag}_{\ga}(t_n^+)a^{I\dag}_{\de}(t_n^+) a^I_{\xi}(t_n)a^I_{\sig}(t_n)O(t_{n+1})\cdots O(t_{k})\right]  |\Phi_0^N \ran  \de_{k,n+m}.
\end{eqnarray} 
\end{widetext}
Note that the factorization of the time ordered product, by inserting a $|\Phi_0^N \ran \lan \Phi_0^N |$, is possible because the Wick theorem normal-orders these products with respect to the reference state, $| \Phi_0^N \ran$. In other words, 
both Eqs.~(\ref{vterm}) and~(\ref{redu}) lead to exactly the same results after all Wick contractions have been carried out.

All possible orders in which 
a general $O(t)$ enters Eq.~(\ref{redu}) are equivalent and are accounted for by 
the binomial factors.
The factor $\binom{3}{1}$ accounts for all the possible ways, eventually decided by contractions, in which the six creation/annihilation operators in $W(t_n)$ can be separated in the two factors [see also Eq.~(\ref{cd_factors}) below].
We also include an additional factor $\binom{3}{1}$ coming from all the possible ways to choose one creation/annihilation operator among the three possible pairs. 
The correct time ordering for creation and annihilation 
operators associated with $W(t_n)$ is preserved using 
$a^{\dag}(t^+_n)$. 

With this decomposition, we can identify the second line of Eq.~(\ref{redu}) as an $m$-th order contribution (with $a$ 2B and $m-a$ 3B operators) to the perturbative expansion of
$G^{4-{\rm pt}}_{\sig\xi,\ga\de}(t_n , t_n ; t_n^+,  t_n^+) = G^{II}_{\sig\xi,\ga\de}(t_n-t_n^+)$. 
Collecting all possible contributions of form (\ref{red_cut}) and (\ref{redu}) 
in which the first $n$ operators are unchanged, the $k$-th order interaction-reducible 
contribution to $G$ becomes:
\begin{widetext}
\begin{eqnarray}
G_{\al\be}^{(k-{\rm th})}(t-t') \to &\left(\frac{-\ii}{\hbar}\right)^{n+1}\frac{1}{n!}\binom{n}{q-a} & \idotsint \textrm{d} t_{n} 
 \lan\Phi_0^N | \T \left[ a^I_{\al}(t) a^{I\dag}_{\be}(t') O(t_1)\cdots O(t_{n-1})a^{I\dag}_{\mu}(t_n^+) a^I_{\ta}(t_n)\right] |\Phi_0^N \ran_{\textrm{int-irr}}
  \nn  \\
&& \quad \quad \times \underbracket{W_{\mu\ga\de,\ta\sig\xi} \,\frac{\ii\, \hbar}{(2!)^2}\, G^{{II}\, (m-{\rm th},\textrm{a})}_{\sig\xi,\ga\de}\,(t_n-t_n^+)}_{\mbox{$U^{\textrm{eff}}_{\mu\ta}$}} \; ,
\label{redu2}
\end{eqnarray} 
\end{widetext}
where $G^{{II}\, (m-{\rm th},\textrm{a})}$ sums all the diagrams at $m$-th order with 
$a$ two-body operators. Note that the last term no longer depends on time and can be seen as an energy-independent correction to the 1B potential. We can automatically take into 
account these interaction-reducible terms by reformulating the initial hamiltonian to include 
the effective 1B vertex:
\begin{equation}
\widetilde{U}_{\mu\ta} \to U_{\mu\ta}+ W_{\mu\ga\de,\ta\sig\xi}\,\frac{\ii\hbar}{(2!)^2}\,\underbracket{G^{II}_{\sig\xi,\ga\de}\,(t- t^+)}_{\mbox{$-\frac{\ii}{\hbar}\,\rho^{2B}_{\sig\xi,\ga\de}$}}
\end{equation}
where now we use an \emph{exact} $G^{II}$. The perturbative expansion obtained 
with this effective interaction should only contain interaction-irreducible diagrams  to avoid double 
counting.
 
Note that in Eq.~(\ref{redu2}) we automatically obtain the correct symmetry factor $1/(2!)^2$ associated with the contraction of $W$ with the two pairs of incoming and outgoing lines 
of $G^{II}$.
In the general case, a $c$-body vertex can be reduced to a $d$-body one (with $d<c$) by 
using a $(c-d)$-body GF.  The overall combinatorial factor in that case will be:
\begin{equation}
\frac{1}{(c!)^2}\left(\frac{c!}{d!(c-d)!}\right)^2=\underbrace{\frac{1}{(d!)^2}}_{\textrm{new vertex}}\,\,\underbrace{\frac{1}{((c-d)!)^2}}_{c-d\, \textrm{equal lines}} \, .
\label{cd_factors}
\end{equation} 
This yields both the correct combinatorial factors entering the new effective $d$-body vertex
and the symmetry factor associated with the contraction with the ($c-d$)-body GF.  
The above arguments can be  generalized to any starting $n$-body Hamiltonian.  
Applying these derivation to all possible cases for a 3B Hamiltonians leads to the effective interactions discussed in Eqs.~(\ref{ueff}) and~(\ref{veff}).

\bibliographystyle{apsrev4-1}
\bibliography{biblio}

\end{document}